\documentclass[lettersize,journal]{IEEEtran}
\usepackage{amsmath,amsfonts,amssymb}
\usepackage{algorithmic}
\usepackage{algorithm}
\usepackage{array}
\usepackage{textcomp}
\usepackage{url}
\usepackage{verbatim}
\usepackage{graphicx}
\usepackage{cite}
\usepackage{mathtools}
\usepackage{amsthm}
\usepackage{cuted}      % for full-width equations in two-column mode
\usepackage{stfloats}   % improves placement of double-column material
\usepackage[caption=false,font=footnotesize]{subfig}
\usepackage{booktabs}
\usepackage{makecell}
\usepackage{xcolor}
\usepackage[colorlinks]{hyperref}

\begin{document}
\title{HARQ-CC-Aided Slow Fluid Antenna Multiple Access with Highly Correlated Ports: An LST-Based Performance Analysis}

\author{Sixu Han, % ~\IEEEmembership{Staff,~IEEE,}
        Kai-Kit Wong,~\IEEEmembership{Fellow,~IEEE},
        Hanjiang Hong,~\IEEEmembership{Member,~IEEE},
        % <-this % stops a space
\vspace{-7mm}

\thanks{The work of K. K. Wong is supported by the Engineering and Physical Sciences Research Council (EPSRC) under Grant EP/W026813/1. The work of H. Shin is supported by the National Research Foundation of Korea (NRF) grant funded by the Korean government (MSIT) (RS-2025-00556064 and RS-2025-25442355), and by the Ministry of Science and ICT (MSIT), Korea, under the ITRC (Information Technology Research Center) support program (IITP-2025-RS-2021-II212046), supervised by the IITP (Institute for Information \& Communications Technology Planning \& Evaluation).}

\thanks{S. Han, K. K. Wong, and H. Hong are with the Department of Electronic and Electrical Engineering, University College London, London, United Kingdom. K. K. Wong is also affiliated with the Department of Electronic Engineering, Kyung Hee University, Yongin-si, Gyeonggi-do 17104, Korea (e-mail: \{sixu.han.22, kai-kit.wong, hanjiang.hong\}@ucl.ac.uk).}

\thanks{Corresponding author: Kai-Kit Wong.}
}

\markboth{Journal of \LaTeX\ Class Files,~Vol.~14, No.~8, August~2021}%
{Shell \MakeLowercase{\textit{et al.}}: A Sample Article Using IEEEtran.cls for IEEE Journals}


\maketitle

\begin{abstract}
Hybrid automatic repeat request with chase combining (HARQ-CC) improves the reliability of slow fluid antenna multiple access (\emph{s}FAMA) through multi-round combining. However, existing analysis has not fully utilized the structure of densely spaced and highly correlated fluid antenna system (FAS) ports to derive tractable per-round characterizations, thereby maintaining a computationally intensive process. This paper re-investigates downlink HARQ-CC-aided \emph{s}FAMA with densely-spaced and highly-correlated FAS configuration. Under a spatial block correlation model, we first formulate two validity-corrected high-correlation approximations for the per-round selected-port signal-to-interference ratio (SIR) distribution and its Laplace--Stieltjes transform (LST): a Marcum-$Q$-kernel route and a lower-complexity step-threshold route. Closed-form expressions are also obtained for block-representative antenna selection (BR-AS) and fixed-position antenna (FPA). Then, the per-round characteristics are used to evaluate the multi-round accumulated-SIR distribution through SIR-domain Stieltjes convolution and numerical LST inversion, yielding the outage probability, average number of transmissions, and payload throughput. Numerical results show close agreement between the two evaluation methods. The analytical FAS results are conservative relative to simulation but preserve the performance trends and receiver ordering. The FAS receiver consistently outperforms the benchmarks, while the payload-throughput gain from increasing the HARQ transmission limit becomes marginal under severe multiuser interference.
\end{abstract}
%\vspace{-2mm}

%\vspace{-2mm}
\section{Introduction}
\label{sec:introduction}
\IEEEPARstart{M}{assive} connectivity represents a fundamental requirement for six-generation (6G) wireless networks, necessitating the simultaneous support of unprecedented numbers of connected devices \cite{intro1,intro2,intro3}. Massive multiple-input multiple-output (MIMO) \cite{MIMO1,MIMO2,MIMO3} improves spectral efficiency and coverage by exploiting large antenna arrays for multiuser spatial multiplexing and beamforming. Non-orthogonal multiple access (NOMA) \cite{NOMA,NOMA1,NOMA2,ding2017survey} and rate-splitting multiple access (RSMA) \cite{RSMA,RSMA1,clerckx2023primer} have also been widely studied for multiuser MIMO, using non-orthogonal signal superposition and message splitting, respectively, to manage multiuser interference. 
Nonetheless, their practical gains typically rely on channel state information (CSI)-dependent transmission design and receiver-side successive interference cancellation (SIC), which may introduce substantial signalling and processing overhead in massive-connectivity scenarios \cite{FAS2}.

Fluid antenna system (FAS) \cite{FAS,FAS1,FAS3,hong2026survey,FAS5} has emerged as a promising physical-layer technology for exploiting spatial channel variations within a compact aperture, where a single radio-frequency (RF) chain can be connected to one of multiple candidate ports to adapt the effective antenna position to the instantaneous propagation and interference conditions \cite{FAS1}.
The concept of FAS was first introduced in 2020 \cite{FAS4}, and has been widely studied in various channels and scenarios\cite{spatialblockmodel,hong2025FASOFDM,finiteblocklengthFAS,11643249,11551155,11594105}. Recent attempts have also studied its channel estimation \cite{10375559, 10751774, 11440310, 11586648, 11555202}.
This spatial diversity of FAS is particularly attractive in interference-limited multiuser networks, where reliable reception depends on both enhancing the desired signal and avoiding strong co-channel interference \cite{FAS2}. Building on FAS, fluid antenna multiple access (FAMA) provides a receiver-side multiple access framework for massive connectivity \cite{FAMA,FAMA1}. FAMA receiver can naturally identify the optimal port with interference nulling, benefiting from the spatial diversity of FAS. Depending on the port-update time scale, FAMA can be broadly categorized into fast FAMA (\emph{f}FAMA) \cite{fFAMA,waqar2026fast} and slow FAMA(\emph{s}FAMA) \cite{sFAMA,sFAMA3,sFAMA2,10663271,SFAMAURA,hong2025multi}. Specifically, \emph{f}FAMA exploits symbol-level spatial fluctuations through rapid port switching to mitigate inter-user interference, whereas \emph{s}FAMA updates the selected port at the channel-coherence or block level. The canonical \emph{s}FAMA selects the port with the most favorable signal-to-interference ratio (SIR) over the available candidate positions.

While spatial port selection can substantially improve the instantaneous reception condition, the reliability offered by a single \emph{s}FAMA transmission remains ultimately limited by the fading and interference state within the corresponding transmission block. This becomes particularly important for 6G services requiring high reliability under dense spectrum reuse. Hybrid automatic repeat request with chase combining (HARQ-CC) \cite{HARQCC,HARQ5G,HARQ1} is a fundamental mechanism for improving link reliability by retransmitting unsuccessfully decoded packets and coherently combining the received replicas. HARQ has also been investigated for NOMA-based
\cite{cai2018performance,wu2022minimizing} and RSMA-based
\cite{abidrabbu2023novel,loli2024hybrid,liu2024performance} multiuser transmission.
From the FAMA perspective, HARQ introduces an additional and complementary source of diversity: whereas \emph{s}FAMA searches for favorable reception conditions across space within each transmission round, HARQ-CC accumulates useful reception opportunities across transmission rounds. The combination of these two mechanisms therefore provides a natural space-time reliability architecture for FAS-enabled massive access. This synergy, namely HARQ-\emph{s}FAMA, was first introduced in \cite{HARQsFAMA}, where each user equipment (UE) re-selects the receive port that maximizes the instantaneous SIR in every HARQ round, such that decoding performance is determined by the accumulated selected-port SIR over multiple rounds. Existing analysis has evaluated this accumulated SIR through direct convolution and demonstrated the resulting multi-round reliability enhancement. 
However, the existing analysis does not exploit the spatial structure induced by densely deployed and highly correlated FAS ports, thereby remaining computationally demanding. The cumulative distribution function (CDF) of the per-round selected-port SIR is expressed as a product of blockwise terms, each involving only a one-dimensional quadrature. Consequently, its multi-round evaluation relies on repeated SIR-domain convolution. This complexity becomes particularly relevant for densely deployed FAS ports, where the correlation is high.

To address this issue, this paper investigates downlink HARQ-CC-aided \emph{s}FAMA and develops a tractable performance analysis framework particularly for densely deployed and highly correlated FAS ports. By combining a spatial block correlation model \cite{spatialblockmodel,finiteblocklengthFAS} with the leading high-correlation Marcum-$Q$ term in~\cite{HARQsFAMA}, we construct an admissible endpoint-corrected kernel CDF and derive its CDF-based LST. We further develop a lower-complexity, validity-corrected step-threshold reduction. Based on these expressions, the accumulated-SIR distribution is evaluated through two computational methods: Stieltjes convolution in the SIR domain and numerical inversion in the LST domain. The main contributions of this paper are summarized as follows:
\begin{itemize}    
    \item We develop a tractable HARQ-CC-aided \emph{s}FAMA framework in densely correlated FAS deployments. First, the selected-port SIR is characterized on a per-round basis by exploiting the high-correlation structure of densely deployed FAS ports. Secondly, these per-round statistics are incorporated into a multi-round HARQ-CC analysis to characterize the accumulated SIR and the resulting transmission performance. This framework provides both SIR-domain and LST-domain routes for analyzing HARQ-CC-aided sFAMA under dense spatial correlation. 

    \item For the per-round analysis, we build on the spatial block-correlation model and leading Marcum-\(Q\) term in \cite{HARQsFAMA} to formulate two admissible high-correlation conditional SIR CDF approximations for the FAS receiver. The first is an endpoint-corrected Marcum-$Q$-kernel form, while the second is a lower-complexity step-threshold form. Endpoint normalization and CDF validity correlations are introduced to ensure the resulting approximations satisfy the required CDF boundary conditions. Based on the corrected CDFs, we derive the corresponding LST representations. Closed-form conditional SIR CDFs and LSTs are also derived for the block-representative antenna selection (BR-AS) and fixed-position antenna (FPA). 
    
    \item Using the derived per-round distributions, the accumulated SIR distribution over multiple HARQ-CC rounds are characterized through two approaches: Stieltjes convolution in SIR-domain and numerical inversion in LST domain. For the latter, we formulate an efficient common-abscissa implementation of the Gaver--Wynn--Rho inversion that reuses the required single-round LST evaluations across HARQ combining orders.

    \item Numerical results demonstrate close agreement between the SIR-domain convolution approach and LST-domain evaluation approach, while Monte Carlo simulations corroborate the predicted performance trends and relative receiver ordering. The FAS receiver consistently outperforms the BR-AS and FPA benchmarks. The results further reveal performance saturation under port densification over a fixed aperture and quantify how the SIR decoding threshold, maximum number of HARQ transmissions, and interferer activity affect the outage probability, retransmission overhead, and payload throughput.
\end{itemize}

\section{System Model}
\label{sec:system_model}
We consider the downlink HARQ-\emph{s}FAMA model, where a base station (BS) with \(N_t=U\) conventional fixed antennas serves \(U\) UEs over the same time--frequency resource. Each UE is equipped with a one-dimensional FAS (1D-FAS) of physical length \(W\lambda\), where \(\lambda\) denotes the carrier wavelength. The FAS consists of \(K\) uniformly spaced candidate ports, among which only one port is connected to a single RF chain in each fading block. The system operates in an interference-limited regime. Consistent with the \emph{s}FAMA protocol, the active receive port is updated once per HARQ round. Each HARQ round occupies one quasi-static fading block of \(L\) channel uses, over which the channel remains constant, while the fading blocks are independent across HARQ rounds. 

For UE~\(u\), the \(q\)-th information packet \(\mathbf d_u^{(q)}\in\{0,1\}^{M_{\mathrm s}}\) is mapped to an \(L\)-symbol coded-modulated block \(\mathbf s_u^{(q)}\in\mathbb C^L\). The initial spectral efficiency is therefore \(R_0\triangleq M_{\mathrm s}/L\) bits/channel use, with the corresponding SIR decoding threshold \(\gamma_{\rm th}=2^{R_0}-1\). Equal-power transmission is assumed for all UEs, and packet-header overhead is neglected. We adopt per-UE packet-level stop-and-wait HARQ-CC, where each UE has at most one active packet under retransmission. Once packet \(\mathbf d_u^{(q)}\) becomes active, one replica of \(\mathbf s_u^{(q)}\) is transmitted in each HARQ round until successful decoding or outage declaration after at most \(C\) rounds. The HARQ round index is denoted by \(i\in\{1,\ldots,C\}\), with \(i=1\) corresponding to the initial transmission.

Since the tagged packet may experience multiple HARQ rounds, we omit the global fading-block index and use the superscript \((i)\) to denote the quantities in the fading block occupied by the \(i\)-th HARQ round. According to the \emph{s}FAMA reception rule, let \(k_u^{(i)}\) denote the receive port selected by UE~\(u\) in round \(i\). The received signal observed at this selected port is
\begin{equation}
\label{eq:ruqi_rx}
\mathbf r_{u,i}^{(q)}
=
g_{(u,u),k_u^{(i)}}^{(i)}\,\mathbf s_u^{(q)}
+
\sum_{\substack{\tilde u=1\\ \tilde u\neq u}}^{U}
a_{\tilde u}^{(i)}
g_{(\tilde u,u),k_u^{(i)}}^{(i)}
\mathbf s_{\tilde u}^{(i)}
+
\boldsymbol\eta_{u,k_u^{(i)}}^{(i)} .
\end{equation}
Here, \(g_{(\tilde u,u),k}^{(i)}\) denotes the effective block-fading channel coefficient of the stream intended for UE~\(\tilde u\), observed at the \(k\)-th port of UE~\(u\) in HARQ round \(i\). The channel coefficients across the \(K\) FAS ports are generally spatially correlated. Accordingly, for For the link associated with the BS stream intended for UE~\(\tilde u\), as observed by UE~\(u\) in HARQ round \(i\), define $\mathbf g_{(\tilde u,u)}^{(i)} \triangleq \big[ g_{(\tilde u,u),1}^{(i)},\ldots, g_{(\tilde u,u),K}^{(i)} \big]^{\mathsf T} \in\mathbb C^K $. The desired block $\mathbf s_u^{(q)}$ is the coded-modulated replica of the tagged packet, whereas $\mathbf s_{\tilde u}^{(i)}\in\mathbb C^L$ denotes the symbol block transmitted for UE~$\tilde u$ in the same fading block, with its packet index not explicitly tracked. The indicator \(a_{\tilde u}^{(i)}\in\{0,1\}\) specifies whether this interfering stream is active on the considered time--frequency resource in round \(i\). The AWGN vector \(\boldsymbol\eta_{u,k_u^{(i)}}^{(i)}\in\mathbb C^L\) at the selected port has i.i.d. entries with zero mean and variance \(\sigma_\eta^2\).

Following the spatial block correlation model in~\cite{spatialblockmodel,finiteblocklengthFAS}, we model $\mathbf g_{(\tilde u,u)}^{(i)} \sim \mathcal{CN}\!\left(\mathbf 0,\sigma^2\widehat{\boldsymbol{\Sigma}}\right)$, where \(\sigma^2\) denotes the average channel-power scaling factor and \(\widehat{\boldsymbol{\Sigma}}\) is the normalized port correlation matrix.
The \(K\) FAS ports are partitioned into \(B\) disjoint blocks \(\{\mathcal K_b\}_{b=1}^{B}\), where \(|\mathcal K_b|=L_b\) and \(\sum_{b=1}^{B}L_b=K\). Let \(b(k)\) denote the block index of port \(k\). The block correlation model approximates the normalized spatial correlation matrix by
\begin{equation}
\label{eq:Sigma_hat_blockdiag}
\widehat{\boldsymbol{\Sigma}}
=
\operatorname{diag}(\mathbf A_1,\ldots,\mathbf A_B)
\in\mathbb R^{K\times K}.
\end{equation}
where each block \(\mathbf A_b\in\mathbb R^{L_b\times L_b}\) is given by
\begin{equation}
\label{eq:Ab_common_form}
\mathbf A_b
=
(1-\mu_b^2)\mathbf I_{L_b}
+
\mu_b^2\mathbf 1_{L_b}\mathbf 1_{L_b}^{\mathsf T}.
\end{equation}
Here, \(\mathbf I_{L_b}\) is the \(L_b\times L_b\) identity matrix, \(\mathbf 1_{L_b}\) is the all-one vector of length \(L_b\), and \(\mu_b^2\) is the within-block correlation coefficient, with \(\mu_b\in(0,1)\). For analytical tractability, we assume a common within-block parameter, i.e., \(\mu_b=\mu\) for all \(b=1,\ldots,B\).
Under this model, the effective channel coefficient associated with the BS stream intended for UE~\(\tilde u\), as observed at the \(k\)-th port of UE~\(u\) in round \(i\), can be represented as
\begin{equation}
\label{eq:common_gauss_multiblock}
\begin{aligned}
g_{(\tilde u,u),k}^{(i)}
={}&
\sigma\Big(
\sqrt{1-\mu^2}\,x_{\tilde u,k}^{(i)}
+\mu x_{\tilde u,b(k)}^{(i)}
\Big)  \\
&\quad
+\mathrm{j}\sigma\Big(
\sqrt{1-\mu^2}\,y_{\tilde u,k}^{(i)}
+\mu y_{\tilde u,b(k)}^{(i)}
\Big),
\end{aligned}
\end{equation}
where all real Gaussian variables are mutually independent with zero mean and variance \(1/2\), independently across users, ports, blocks, and HARQ rounds.
For notational compactness, let \(\chi_\mu \triangleq \mu/\sqrt{1-\mu^2}\). The normalized channel power at the \(k\)-th port is then defined as
\begin{equation}
\label{eq:S_tilde_u_k}
S_{\tilde u,k}^{(i)}
\triangleq
\bigl(x_{\tilde u,k}^{(i)}
+\chi_\mu x_{\tilde u,b(k)}^{(i)}\bigr)^2
+
\bigl(y_{\tilde u,k}^{(i)}
+\chi_\mu y_{\tilde u,b(k)}^{(i)}\bigr)^2 .
\end{equation}
Then,
\begin{equation}
\label{eq:g_power_normalized}
|g_{(\tilde u,u),k}^{(i)}|^2
=
\sigma^2(1-\mu^2)S_{\tilde u,k}^{(i)} .
\end{equation}
Under the \emph{s}FAMA protocol, the active port is updated once per HARQ round according to the instantaneous channel realization and activity pattern.
Specifically, in round \(i\), the selected port is given by
\begin{equation}\label{eq:ku_i_HARQ_compact}
k_u^{(i)}=\arg\max_{k}\frac{|g_{(u,u),k}^{(i)}|^2}{\sum_{\substack{\tilde u=1\\ \tilde u\neq u}}^{U}a_{\tilde u}^{(i)} |g_{(\tilde u,u),k}^{(i)}|^2 }=\arg\max_{k}\frac{S_{u,k}^{(i)}}{I_{u,k}^{(i)}},
\end{equation}
where $S_{u,k}^{(i)}$ follows from \eqref{eq:S_tilde_u_k} by setting \(\tilde u=u\), and $I_{u,k}^{(i)}$ is now defined as
\begin{equation}\label{eq:I_u_k_i_sum_short}
I_{u,k}^{(i)}
\triangleq
\sum_{\substack{\tilde u=1\\ \tilde u\neq u}}^{U}
a_{\tilde u}^{(i)}\, S_{\tilde u,k}^{(i)}.
\end{equation}

We define the number of active interferers in round \(i\) as $A^{(i)}
\triangleq
\sum_{\substack{\tilde u=1\\ \tilde u\neq u}}^{U}
a_{\tilde u}^{(i)}$.
Following the interference-limited activity formulation in~\cite{HARQsFAMA}, we evaluate the per-round SIR distribution over the nonzero-interference rounds. Accordingly, the zero-truncated activity count is defined as
$\bar A^{(i)}
\triangleq
A^{(i)}\,\big|\,\bigl(A^{(i)}\ge 1\bigr)$. 
Under the Bernoulli activity model, the indicators
$\{a_{\tilde u}^{(i)}\}$ are i.i.d. across users and HARQ rounds, with
$\mathbb P(a_{\tilde u}^{(i)}=1)=p_{\rm a}$. The activity probability
$p_{\rm a}$ is treated as an exogenous system parameter and remains fixed
throughout the HARQ process, whereas the activity realizations vary
independently from round to round. Hence, the distribution of
$\bar A^{(i)}$ does not depend on $i$, and we omit the round index when
analyzing a generic HARQ round. For $m=1,\ldots,U-1$, the probability mass
function of $\bar A$ is
\begin{equation}
\label{eq:Abar_pmf}
\mathbb P\bigl(\bar A=m\bigr)
=
\binom{U-1}{m}
\frac{
p_{\rm a}^{m}(1-p_{\rm a})^{U-1-m}
}{
1-(1-p_{\rm a})^{U-1}
}.
\end{equation}

\section{Per-Round SIR Characterization and LST Analysis for HARQ-CC-aided \emph{s}FAMA}
\label{sec:per_round_lst}
\subsection{Conditional Per-Round SIR CDFs}
Under the adopted independent block fading and independent activity
realizations across HARQ rounds, the per-round SIRs are independent and
identically distributed for a fixed activity probability \(p_{\rm a}\).
The round index is therefore omitted. For a generic receiver architecture
\(\mathsf{R}\in\{\text{FAS},\text{BR-AS},\text{FPA}\}\), let
$
F_m^{\mathsf R}(\gamma)
\triangleq
F_{\mathrm{SIR}_{u,\mathsf R}}
\bigl(\gamma\mid \bar A=m\bigr)
$
denote the conditional CDF of the per-round SIR under receiver
architecture \(\mathsf R\), given \(\bar A=m\), where
\(m\in\{1,\ldots,U-1\}\).
We are primarily focusing on the \emph{s}FAMA case with the FAS receiver, i.e., ${\mathsf R} = {\text{FAS}}$. The other two cases are also analyzed as benchmarks. 
For the FAS receiver, we consider two tractable high-correlation CDF approximations: the first retains the leading Marcum-$Q$ kernel, whereas the second applies an additional step-threshold reduction. The raw
Marcum-$Q$ curve is endpoint-normalized to remove its spurious atom at the
origin, whereas the raw threshold curve is corrected for its zero endpoint,
range, and monotonicity. Only the resulting curves are supplied to the HARQ
convolution or LST calculations. These corrections enforce the properties
required by the continuous-fading model; they do not establish an accuracy
bound for the underlying high-correlation approximations.

To distinguish the leading generalized Marcum-$Q$ term in the
single-port conditional CDF factor $\Xi_m$ appearing in
\cite[eq.~(35)]{HARQsFAMA}, define
\begin{equation}
\label{eq:mq_kernel_definition}
\mathcal Q_m(\gamma;r,\tilde r)
\triangleq
Q_m\!\left(
\sqrt{\frac{2\chi_\mu^2\gamma\tilde r}{1+\gamma}},
\sqrt{\frac{2\chi_\mu^2r}{1+\gamma}}
\right),
\qquad \gamma\geq0,
\end{equation}
where $Q_m(\cdot,\cdot)$ denotes the generalized Marcum-$Q$
function. Under the adopted block-correlation model,
$\Xi_m$ admits the decomposition
\begin{equation}
\label{eq:Xi_complete_relation}
\Xi_m(\gamma;r,\tilde r)
=
\mathcal Q_m(\gamma;r,\tilde r)
-
\mathcal R_m(\gamma;r,\tilde r),
\end{equation}
where $\mathcal R_m(\gamma;r,\tilde r)\geq0$ denotes the finite
double-sum term in~\cite[eq.~(35)]{HARQsFAMA}.
Although $\mathcal Q_m$ alone does not equal $\Xi_m$ for a fixed
$\mu<1$, the high-correlation limit $\mu\to1$ yields the blockwise
approximation
\begin{equation}
\label{eq:Xi_power_hc_approx}
\bigl[\Xi_m(\gamma;r,\tilde r)\bigr]^{L_b}
\approx
\bigl[\mathcal Q_m(\gamma;r,\tilde r)\bigr]^{L_b}.
\end{equation}
Using the right-hand side of
\eqref{eq:Xi_power_hc_approx} in the blockwise CDF construction
yields the following raw Marcum-$Q$ approximation to the conditional
per-round SIR CDF:
\begin{equation}
\label{eq:FSIR_MQ_raw_integral}
\begin{aligned}
\widehat F_{m,\mathrm{mq}}^{\mathrm{FAS}}(\gamma)
&\triangleq
\prod_{b=1}^{B}
\Bigg[
\frac{1}{\Gamma(m)}
\int_{0}^{\infty}\!\!\int_{0}^{\infty}
\tilde r^{\,m-1} e^{-(r+\tilde r)}
\\[-1mm]
&\qquad\qquad\times
\bigl[\mathcal Q_m(\gamma;r,\tilde r)\bigr]^{L_b}
\,dr\,d\tilde r
\Bigg].
\end{aligned}
\end{equation}
Using the Gauss--Laguerre rules and weights, this becomes
\begin{equation}
\label{eq:FSIR_MQ_raw_GLQ}
\begin{aligned}
\widehat F_{m,\mathrm{mq}}^{\mathrm{FAS}}(\gamma)
&\approx
\prod_{b=1}^{B}
\Bigg[
\frac{1}{\Gamma(m)}
\sum_{\ell=1}^{N_r}
\sum_{\tilde\ell=1}^{N_{\tilde r}}
w_{\ell}\,
\widetilde w_{\tilde\ell}^{(m)}
\\[-1mm]
&\qquad\times
\Bigl[
\mathcal Q_m\!\left(
\gamma;
\zeta_{\ell},
\widetilde\zeta_{\tilde\ell}^{(m)}
\right)
\Bigr]^{L_b}
\Bigg].
\end{aligned}
\end{equation}
Here, $\zeta_\ell$ and $\widetilde\zeta_{\tilde\ell}^{(m)}$ are the roots
of $L_{N_r}(x)$ and $L_{N_{\tilde r}}^{(m-1)}(x)$, respectively, and
their weights are
\begin{align}
w_\ell
&\triangleq
\frac{\zeta_\ell}{(N_r+1)^2\bigl[L_{N_r+1}(\zeta_\ell)\bigr]^2},
\label{eq:GL_w_r}\\
\widetilde w_{\tilde\ell}^{(m)}
&\triangleq
\frac{\Gamma(N_{\tilde r}+m)}{N_{\tilde r}!\,(N_{\tilde r}+1)^2}\,
\frac{\widetilde\zeta_{\tilde\ell}^{(m)}}{\bigl[L_{N_{\tilde r}+1}^{(m-1)}(\widetilde\zeta_{\tilde\ell}^{(m)})\bigr]^2}.
\label{eq:GL_w_tilder}
\end{align}
As $L_b$ increases, this exponentiation sharpens the
transition of the Marcum-$Q$ term. In the considered dense
high-correlation regime, we therefore approximate the per-block
selection term by a step function located at its maximum-descent point,
with threshold $\delta_{m,b}(\tilde r_{u,b};\gamma)$. For $\gamma>0$, the resulting raw
threshold approximation is
\begin{equation}
\label{eq:FSIR_threshold_raw_integral}
\begin{aligned}
\widehat F_{m,\mathrm{th}}^{\mathrm{FAS}}(\gamma)
&\triangleq
\prod_{b=1}^{B}
\Bigg[
1-\frac{1}{\Gamma(m)}
\int_{0}^{\infty}
\tilde r_{u,b}^{\,m-1}
\\[-1mm]
&\quad\times
\exp\!\left(
-\tilde r_{u,b}
-\delta_{m,b}(\tilde r_{u,b};\gamma)
\right)
\mathrm d\tilde r_{u,b}
\Bigg].
\end{aligned}
\end{equation}
where $\delta_{m,b}(\tilde r_{u,b};\gamma)$ is given by
\begin{equation}
\label{eq:delta_36_like_final_correct}
\begin{aligned}
\delta_{m,b}(\tilde r_{u,b};\gamma)
&=
\Bigg(
\sqrt{\gamma \tilde r_{u,b}}
+
\beta\,
\frac{
\bigl(m-\frac{1}{2}\bigr)\beta
-\kappa_b\sqrt{\gamma \tilde r_{u,b}}
}{
\kappa_b\bigl(m-\frac{1}{2}\bigr)\beta
+\sqrt{\gamma \tilde r_{u,b}}
}
\Bigg)^{\!2},
\\[-1mm]
\kappa_b
&=
\sqrt{\frac{L_b-1}{2\pi}},
\qquad
\beta
=
\frac{1}{\chi_\mu}
\sqrt{\frac{\gamma+1}{2}} .
\end{aligned}
\end{equation}
The step-threshold expression is used for \(L_b>1\), as satisfied by all
dense-port configurations considered in this work.
Using Gauss-Laguerre quadrature again, we obtain
\begin{equation}
\label{eq:FSIR_threshold_raw_GLQ}
\begin{aligned}
\widehat F_{m,\mathrm{th}}^{\mathrm{FAS}}(\gamma)
&\approx
\prod_{b=1}^{B}
\Bigg[
1-\frac{1}{\Gamma(m)}
\sum_{\tilde\ell=1}^{N_{\tilde r}}
\widetilde w_{\tilde\ell}^{(m)}
\\[-1mm]
&\qquad\times
\exp\!\left[
-\delta_{m,b}\!\left(
\widetilde\zeta_{\tilde\ell}^{(m)};\gamma
\right)
\right]
\Bigg].
\end{aligned}
\end{equation}
where \(\widetilde\zeta_{\tilde\ell}^{(m)}\) is the corresponding
quadrature node, and \(\widetilde w_{\tilde\ell}^{(m)}\) is given
in~\eqref{eq:GL_w_tilder}.

To connect the step-like approximation with a simpler antenna-selection reference, we introduce a BR-AS benchmark, where the dense ports inside each correlation block are not exploited. Specifically, one representative fixed-position antenna element is retained from each correlation block, resulting in \(B\) representative receive branches, among which the branch with the largest instantaneous SIR is selected.
For the representative element in the \(b\)-th block, the success condition \(\mathrm{SIR}_{u,b}>\gamma\) is equivalent to \(r_{u,b}>\gamma\tilde r_{u,b}\). Thus, the threshold reduces to \(\delta_{m,b}(\tilde r_{u,b};\gamma)=\gamma\tilde r_{u,b}\). Substituting this into the integral structure in \eqref{eq:FSIR_threshold_raw_integral}, the per-branch success probability is given by
\begin{equation}
\label{eq:gamma_integral_eval}
\frac{1}{\Gamma(m)}
\int_{0}^{\infty}
\tilde r_{u,b}^{\,m-1}
e^{-(1+\gamma)\tilde r_{u,b}}
\,{\rm d}\tilde r_{u,b}
=
\frac{1}{(1+\gamma)^m},
\end{equation}
where the equality follows from the standard Gamma-integral identity. Therefore, for \(\gamma\ge0\), substituting \eqref{eq:gamma_integral_eval} into the same product structure yields the conditional SIR CDF of the BR-AS benchmark as
\begin{equation}
\label{eq:FSIR_BR_conditional}
\begin{aligned}
F_m^{\text{BR-AS}}(\gamma)
&=
\left[1-(1+\gamma)^{-m}\right]^B
\\[-1mm]
&=
1+
\sum_{k=1}^{B}
\binom{B}{k}
(-1)^k
(1+\gamma)^{-mk}.
\end{aligned}
\end{equation}
The BR-AS benchmark captures the block-level selection gain by modeling the \(B\) representative branches as independent, while excluding the additional dense within-block port-sampling gain of FAS.

The conventional FPA receiver is obtained as the single-branch case of this reference. Specifically, when \(B=1\), each UE is equipped with only one fixed receive antenna and no receive-branch selection is performed. Conditioned on \(\bar A=m\), the corresponding conditional SIR CDF is
\begin{equation}
\label{eq:FSIR_FPA_conditional}
F_m^{\text{FPA}}(\gamma)
=
1-(1+\gamma)^{-m},
\qquad \gamma\ge0 .
\end{equation}
Therefore, comparing \eqref{eq:FSIR_BR_conditional} with
\eqref{eq:FSIR_FPA_conditional} quantifies the conventional independent
branch-selection gain, while comparing either
\eqref{eq:FSIR_MQ_corrected} or \eqref{eq:FSIR_threshold_corrected} with
\eqref{eq:FSIR_BR_conditional} identifies the additional gain induced by
dense within-block port sampling.

\subsection{FAS Endpoint Normalization and CDF Validity Correction}
The raw CDF approximations associated with both FAS routes exhibit a spurious positive residual at the zero-SIR endpoint. We next characterize the corresponding residuals and introduce a common endpoint normalization.
For the Marcum-\(Q\) route, integer \(m\geq 1\) gives
\begin{equation}
\label{eq:mq_kernel_zero_endpoint}
\begin{aligned}
\mathcal Q_m(0;r,\tilde r)
&=
Q_m\!\left(0,\sqrt{2\chi_\mu^2 r}\right)
=
\frac{\Gamma\!\left(m,\chi_\mu^2 r\right)}{\Gamma(m)}
\\[-1mm]
&=
e^{-\chi_\mu^2 r}
\sum_{n=0}^{m-1}
\frac{\left(\chi_\mu^2 r\right)^n}{n!}.
\end{aligned}
\end{equation}
where \(\Gamma(\cdot,\cdot)\) denotes the upper incomplete gamma
function. Consequently, the raw Marcum-\(Q\)-kernel curve has the
positive endpoint residual
\begin{equation}
\label{eq:MQ_endpoint_residual}
\begin{aligned}
\rho_{m,\mathrm{mq}}
&\triangleq
\widehat F_{m,\mathrm{mq}}^{\mathrm{FAS}}(0)
\\
&=
\prod_{b=1}^{B}
\int_{0}^{\infty}
e^{-r}
\left[
e^{-\chi_\mu^2 r}
\sum_{n=0}^{m-1}
\frac{(\chi_\mu^2 r)^n}{n!}
\right]^{L_b} dr
\\
&\approx
\prod_{b=1}^{B}
\sum_{\ell=1}^{N_r}
w_{\ell}
\left[
e^{-\chi_\mu^2\zeta_{\ell}}
\sum_{n=0}^{m-1}
\frac{(\chi_\mu^2\zeta_{\ell})^n}{n!}
\right]^{L_b}.
\end{aligned}
\end{equation}
where
\(
0<\rho_{m,\mathrm{mq}}<1
\)
for \(0<\mu<1\). 

For the threshold route, the large-argument approximation used to
obtain the raw threshold curve is likewise non-uniform as the SIR
argument approaches zero from the right. Assuming \(L_b>1\) for all
blocks,
\begin{equation}
\label{eq:threshold_delta_zero_limit}
\lim_{\gamma\to 0^{+}}
\delta_{m,b}
\!\left(
\tilde r_{u,b};\gamma
\right)
=
\frac{
\pi(1-\mu^2)
}{
\mu^2(L_b-1)
},
\end{equation}
independently of \(\tilde r_{u,b}\). Hence,
\begin{equation}
\label{eq:threshold_endpoint_residual}
\rho_{0,\mathrm{th}}
\triangleq
\lim_{\gamma\to 0^{+}}
\widehat F_{m,\mathrm{th}}^{\mathrm{FAS}}(\gamma)
=
\prod_{b=1}^{B}
\left[
1-
\exp\!\left(
-\frac{
\pi(1-\mu^2)
}{
\mu^2(L_b-1)
}
\right)
\right].
\end{equation}
The threshold residual is independent of \(m\) and satisfies
\(
0<\rho_{0,\mathrm{th}}<1
\)
for \(0<\mu<1\).
Both endpoint residuals are removed using the normalization
\begin{equation}
\label{eq:endpoint_affine_mapping}
\mathcal T_{\rho}(x)
\triangleq
\frac{x-\rho}{1-\rho},
\qquad
0<\rho<1.
\end{equation}
This mapping sends the raw endpoint levels \(\rho\) and \(1\) to
\(0\) and \(1\), respectively, while preserving the ordering of the
raw curve.

For the Marcum-\(Q\) route, the endpoint-normalized CDF is therefore
defined as
\begin{equation}
\label{eq:FSIR_MQ_corrected}
\begin{aligned}
F_{m,\mathrm{mq}}^{\mathrm{FAS}}(\gamma)
&\triangleq
\mathcal T_{\rho_{m,\mathrm{mq}}}
\!\left(
\widehat F_{m,\mathrm{mq}}^{\mathrm{FAS}}(\gamma)
\right)
\\[-1mm]
&=
\frac{
\widehat F_{m,\mathrm{mq}}^{\mathrm{FAS}}(\gamma)
-\rho_{m,\mathrm{mq}}
}{
1-\rho_{m,\mathrm{mq}}
},
\qquad \gamma\geq 0.
\end{aligned}
\end{equation}
Equation~\eqref{eq:FSIR_MQ_corrected} removes the spurious zero-SIR
residual while preserving the ordering and unit upper endpoint of the
raw curve. In the numerical implementation, the quadrature expression
in~\eqref{eq:MQ_endpoint_residual} is used for
\(\rho_{m,\mathrm{mq}}\), together with the same outer quadrature as
in the raw-CDF evaluation, thereby enforcing an exact zero endpoint
for the discretized curve.

For the threshold route, endpoint normalization alone is insufficient
because the raw curve may additionally exhibit small range violations
and a local loss of monotonicity in the low-SIR region. We first define
the endpoint-normalized and range-clipped curve as
\begin{equation}
\label{eq:threshold_endpoint_normalized}
\begin{aligned}
\overline F_{m,\mathrm{th}}^{\mathrm{FAS}}(\gamma)
&\triangleq
\left[
\mathcal T_{\rho_{0,\mathrm{th}}}
\!\left(
\widehat F_{m,\mathrm{th}}^{\mathrm{FAS}}(\gamma)
\right)
\right]_{0}^{1}
\\[-1mm]
&=
\left[
\frac{
\widehat F_{m,\mathrm{th}}^{\mathrm{FAS}}(\gamma)
-\rho_{0,\mathrm{th}}
}{
1-\rho_{0,\mathrm{th}}
}
\right]_{0}^{1},
\qquad \gamma>0.
\end{aligned}
\end{equation}
where
$[x]_{0}^{1}\triangleq\min\{1,\max\{0,x\}\}$, and set
$\overline F_{m,\mathrm{th}}^{\mathrm{FAS}}(0)=0$.
The clipping in~\eqref{eq:threshold_endpoint_normalized} enforces the
range constraint, whereas monotonicity is restored by the
running-supremum envelope
\begin{equation}
\label{eq:FSIR_threshold_corrected}
F_{m,\mathrm{th}}^{\mathrm{FAS}}(\gamma)
\triangleq
\sup_{0\leq t\leq\gamma}
\overline F_{m,\mathrm{th}}^{\mathrm{FAS}}(t),
\qquad
\gamma\geq0.
\end{equation}
Since the
resulting curve starts from zero, remains in \([0,1]\), and converges
to one, it is an admissible CDF approximation for both Stieltjes
convolution and CDF-based LST evaluation.
These corrections enforce only the defining CDF properties. They do
not restore the omitted \(\mathcal R_m\) term in the Marcum-\(Q\)
route or remove the modeling error of the threshold approximation.

\subsection{LST-Domain HARQ-CC Analysis}
\label{subsec:lst_outage_evaluation}
For a generic receiver architecture \(\mathsf{R}\), averaging over \(\bar A\) gives the unconditional per-round SIR CDF as
\begin{equation}
\label{eq:FSIR_unconditional_letter}
F_{\mathrm{SIR}_u}^{\mathsf R}(\gamma)
=
\sum_{m=1}^{U-1}
\mathbb P(\bar A=m)
F_m^{\mathsf R}(\gamma).
\end{equation}
When \(\mathsf R=\mathrm{FAS}\), the route index is suppressed in this generic notation, where
\(F_m^{\mathsf R}\) denotes either
\(F_{m,\mathrm{mq}}^{\mathrm{FAS}}\), defined in
\eqref{eq:FSIR_MQ_corrected}, or
\(F_{m,\mathrm{th}}^{\mathrm{FAS}}\), defined in
\eqref{eq:FSIR_threshold_corrected}.
For a fixed activity state \(m\),
the corresponding conditional LST is defined as
\begin{equation}
\label{eq:LST_conditional_m_def}
\begin{aligned}
\mathcal L_m^{\mathsf R}(s)
&\triangleq
\mathbb E\!\left[
e^{-s\mathrm{SIR}_{u,\mathsf R}}
\,\middle|\,
\bar A=m
\right]
=
\int_{0}^{\infty}
e^{-s\gamma}\,
dF_m^{\mathsf R}(\gamma) \\
&=
s
\int_{0}^{\infty}
e^{-s\gamma}
F_m^{\mathsf R}(\gamma)
\,d\gamma 
\qquad s>0 .
\end{aligned}
\end{equation}
The last equality follows from integration by parts, using the fact that
the per-round SIR is nonnegative and has no atom at the origin under the
considered continuous fading model. The two corrected FAS CDFs enforce
this zero-endpoint property explicitly.

We first consider the BR-AS benchmark. For \(m=1,\ldots,U-1\) and \(s>0\), substituting \eqref{eq:FSIR_BR_conditional} into the CDF-based LST representation in \eqref{eq:LST_conditional_m_def}, we obtain
\begin{align}
\mathcal L_m^{\text{BR-AS}}(s)
&=
1+
\sum_{k=1}^{B}
\binom{B}{k}
(-1)^k
s
\int_{0}^{\infty}
e^{-s\gamma}
(1+\gamma)^{-mk}
\,d\gamma
\nonumber\\
&=
1
+
e^{s}
\sum_{k=1}^{B}
\binom{B}{k}
(-1)^k
s^{mk}
\Gamma(1-mk,s),
\label{eq:LST_BR_conditional_closed}
\end{align}
where \(\Gamma(\cdot,\cdot)\) is the upper incomplete Gamma function.
For the FPA receiver, which can be regarded as the \(B=1\) special case of the BR-AS benchmark, \eqref{eq:LST_BR_conditional_closed} reduces to
\begin{equation}
\mathcal L_m^{\text{FPA}}(s)
=
1
-
e^{s}
s^{m}
\Gamma(1-m,s).
\label{eq:LST_FPA_conditional_closed}
\end{equation}

For the FAS receiver under \emph{s}FAMA, let
\(\nu\in\{\mathrm{mq},\mathrm{th}\}\) identify the validity-corrected
Marcum-$Q$-kernel or step-threshold route, respectively. With the change
of variable \(x=s\gamma\), the conditional LST can be written as
\begin{equation}
\label{eq:LST_FAS_change_variable}
\mathcal L_{m,\nu}^{\text{FAS}}(s)
=
\int_{0}^{\infty}
e^{-x}
F_{m,\nu}^{\text{FAS}}\!\left(\frac{x}{s}\right)
\,{\rm d}x .
\end{equation}
For \(m\in\{1,\ldots,U-1\}\) and \(s>0\), applying an
\(N_\gamma\)-point Gauss--Laguerre rule to the outer integral gives the
validity-corrected Marcum-$Q$-kernel LST
\begin{equation}
\label{eq:LST_FAS_MQ_corrected_GLQ}
\mathcal L_{m,\mathrm{mq}}^{\text{FAS}}(s)
\approx
\sum_{q=1}^{N_\gamma}
\omega_q
F_{m,\mathrm{mq}}^{\text{FAS}}\!\left(
\frac{\xi_q}{s}
\right).
\end{equation}
Here, \(\xi_q\) and \(\omega_q\) denote the nodes and weights of the
\(N_\gamma\)-point Gauss--Laguerre rule for the outer LST integral,
with \(\omega_q\) given by~\eqref{eq:GL_w_r} after replacing
\(N_r\), \(\zeta_\ell\), and \(w_\ell\) with
\(N_\gamma\), \(\xi_q\), and \(\omega_q\), respectively.
Alternatively, substituting the validity-corrected threshold CDF
in~\eqref{eq:FSIR_threshold_corrected} into
\eqref{eq:LST_FAS_change_variable} gives the lower-complexity
approximation
\begin{equation}
\label{eq:LST_FAS_threshold_corrected_GLQ}
\mathcal L_{m,\mathrm{th}}^{\text{FAS}}(s)
\approx
\sum_{q=1}^{N_\gamma}
\omega_q
F_{m,\mathrm{th}}^{\text{FAS}}\!\left(
\frac{\xi_q}{s}
\right).
\end{equation}
Thus, neither raw curve enters the transform calculation directly: the
Marcum-$Q$ route uses the endpoint normalization in
\eqref{eq:FSIR_MQ_corrected}, whereas the threshold route uses the
endpoint-normalized monotone envelope in
\eqref{eq:FSIR_threshold_corrected}.

Finally, averaging the conditional LSTs over the zero-truncated activity
distribution gives
\begin{equation}
\label{eq:LST_unconditional_from_conditional}
\mathcal L_{\mathrm{SIR}_u}^{\mathsf R}(s)
=
\sum_{m=1}^{U-1}
\mathbb P\bigl(\bar A=m\bigr)
\mathcal L_m^{\mathsf R}(s),
\qquad
s>0 .
\end{equation}
For \(\mathsf R=\mathrm{FAS}\), the suppressed route index in
\eqref{eq:LST_unconditional_from_conditional} is again either
\(\mathrm{mq}\) or \(\mathrm{th}\), consistently with the CDF route
selected in~\eqref{eq:FSIR_unconditional_letter}.

\section{Multi-round HARQ-CC Performance Evaluation for \emph{s}FAMA}
\label{sec:lst_harq_outage}
\subsection{HARQ-CC SIR Accumulation and Outage Evaluation}
\label{subsec:accumulated_sir}
For the HARQ-CC analysis, let
\(\mathrm{SIR}_{u,\mathsf R}^{(i)}\) denote the post-reception SIR of UE
\(u\) in the \(i\)-th HARQ round. Owing to the block-fading model and the
across-round i.i.d. activity randomization,
\(\{\mathrm{SIR}_{u,\mathsf R}^{(i)}\}_{i\ge1}\) are i.i.d. copies of the
generic per-round SIR \(\mathrm{SIR}_{u,\mathsf R}\). Hence, the accumulated
SIR after \(j\) HARQ-CC rounds is
\begin{equation}
\label{eq:Gamma_sum_HARQ_CC}
\Gamma_{u,\mathsf R}^{(q)}(j)
=
\sum_{i=1}^{j}
\mathrm{SIR}_{u,\mathsf R}^{(i)},
\qquad
j=1,\ldots,C .
\end{equation}
Successful decoding after \(j\) rounds occurs when
\(\Gamma_{u,\mathsf R}^{(q)}(j)\ge\gamma_{\rm th}\). 
Because both validity-corrected FAS routes provide admissible per-round
CDFs, the accumulated-SIR CDF can be written as the \(j\)-fold Stieltjes
convolution of the per-round CDF, i.e.,
\begin{equation}
\label{eq:F_Gamma_j_main}
F_{\Gamma_{u,\mathsf R}^{(q)}(j)}(\gamma)
=
\left(
F_{\mathrm{SIR}_u}^{\mathsf R}
\right)^{\circledast j}
(\gamma),
\qquad
\gamma\ge 0 .
\end{equation}
Therefore, the
HARQ-CC outage probability after at most \(C\) rounds is
\begin{equation}
\label{eq:Pout_system_def}
\begin{aligned}
P_{\rm out}^{\mathsf R}
&=
\Pr\left\{
\Gamma_{u,\mathsf R}^{(q)}(C)
<
\gamma_{\rm th}
\right\}  \\
&=
F_{\Gamma_{u,\mathsf R}^{(q)}(C)}(\gamma_{\rm th})
=
\left(
F_{\mathrm{SIR}_u}^{\mathsf R}
\right)^{\circledast C}
(\gamma_{\rm th}) .
\end{aligned}
\end{equation}
Since the Stieltjes convolution is converted into multiplication in the LST
domain, the accumulated-SIR LST is given by
\begin{equation}
\label{eq:LST_accumulated_SIR}
\mathcal L_{\Gamma_{u,\mathsf R}^{(q)}(j)}(s)
=
\left(
\mathcal L_{\mathrm{SIR}_u}^{\mathsf R}(s)
\right)^j,
\qquad
j=1,\ldots,C .
\end{equation}
For numerical inversion, define
\begin{equation}
\label{eq:G_j_R_def}
G_{j,\mathsf R}(s)
\triangleq
\frac{
\left(
\mathcal L_{\mathrm{SIR}_u}^{\mathsf R}(s)
\right)^j
}{s},
\qquad s>0 .
\end{equation}
Accordingly, the accumulated-SIR CDF can be recovered from
\begin{equation}
\label{eq:CDF_accumulated_G_def}
F_{\Gamma_{u,\mathsf R}^{(q)}(j)}(\gamma)
=
\mathcal L^{-1}
\left\{
G_{j,\mathsf R}(s)
\right\}(\gamma),
\qquad
\gamma\ge 0 .
\end{equation}
where \(\mathcal L^{-1}\{\cdot\}\) denotes inverse Laplace transformation.
Thus, the \(j\)-round accumulated-SIR CDF can be obtained by numerically
inverting a scalar Laplace-domain function instead of explicitly computing
the \(j\)-fold Stieltjes convolution.

\subsection{Common-Abscissa Evaluation and Gaver--Wynn--Rho Inversion}
\label{subsec:common_abscissa_GWR}
For an inversion point \(\gamma>0\), define
\begin{equation}
\label{eq:common_abscissa_scale}
a_{\gamma}
\triangleq
\frac{\ln 2}{\gamma},
\qquad
s_k
\triangleq
k a_{\gamma},
\quad k=1,\ldots,2M,
\end{equation}
where \(M\) denotes the order of the numerical inverse-Laplace
procedure. Applying the same CDF-based LST identity as in
\eqref{eq:LST_conditional_m_def} to the unconditional CDF in
\eqref{eq:FSIR_unconditional_letter}, followed by the change of variable
\(x=a_{\gamma}\gamma'\), yields
\begin{align}
\mathcal L_{\mathrm{SIR}_u}^{\mathsf R}(s_k)
&=
k\int_{0}^{\infty}
e^{-kx}
F_{\mathrm{SIR}_u}^{\mathsf R}
\left(
\frac{x}{a_{\gamma}}
\right)
\,\mathrm dx
\nonumber\\
&=
k\int_{0}^{\infty}
e^{-x}e^{-(k-1)x}
F_{\mathrm{SIR}_u}^{\mathsf R}
\left(
\frac{x}{a_{\gamma}}
\right)
\,\mathrm dx .
\label{eq:LST_common_abscissa_integral}
\end{align}
Applying an \(N_{\gamma}\)-point Gauss--Laguerre rule to
\eqref{eq:LST_common_abscissa_integral} gives
\begin{equation}
\label{eq:LST_common_abscissa_GLQ}
\mathcal L_{\mathrm{SIR}_u}^{\mathsf R}(s_k)
\approx
k\sum_{q=1}^{N_{\gamma}}
\omega_q
e^{-(k-1)\xi_q}
F_{\mathrm{SIR}_u}^{\mathsf R}
\left(
\frac{\xi_q}{a_{\gamma}}
\right),
\end{equation}
where \(\xi_q\) and \(\omega_q\) denote the node and weight of the
ordinary Gauss--Laguerre rule, respectively. 
Unlike the directly scaled rule
\[
\mathcal L_{\mathrm{SIR}_u}^{\mathsf R}(s_k)
\approx
\sum_{q=1}^{N_\gamma}
\omega_q
F_{\mathrm{SIR}_u}^{\mathsf R}
\left(\frac{\xi_q}{s_k}\right),
\]
which uses \(k\)-dependent CDF arguments,
\eqref{eq:LST_common_abscissa_GLQ} reuses the common set
\(\{\xi_q/a_\gamma\}_{q=1}^{N_\gamma}\) for all
\(k=1,\ldots,2M\). This reduces the number of per-round FAS CDF
evaluations from \(2MN_\gamma\) to \(N_\gamma\) per inversion point.
The two formulations are equivalent before quadrature but may differ
under a finite-order Gauss--Laguerre rule.

Combining \eqref{eq:LST_common_abscissa_GLQ} and
\eqref{eq:G_j_R_def}, the required real-axis transform samples
are evaluated as
\begin{equation}
\label{eq:G_j_common_abscissa_samples}
G_{j,\mathsf R}(s_k)
\approx
\frac{1}{k a_{\gamma}}
\left[
k
\sum_{q=1}^{N_{\gamma}}
\omega_q
e^{-(k-1)\xi_q}
F_{\mathrm{SIR}_u}^{\mathsf R}
\left(
\frac{\xi_q}{a_{\gamma}}
\right)
\right]^j .
\end{equation}
The inverse Laplace transform is evaluated using the
Gaver--Wynn--Rho (GWR) algorithm \cite{GWR}. For \(n=1,\ldots,M\), the \(n\)th
Gaver functional constructed from the \(2M\) samples
\(\{G_{j,\mathsf R}(s_k)\}_{k=1}^{2M}\) is defined as
\begin{align}
\Phi_n^{(j,\mathsf R)}(\gamma)
&\triangleq
\frac{n\ln 2}{\gamma}
\binom{2n}{n}
\sum_{r=0}^{n}
(-1)^r
\binom{n}{r}
\nonumber\\[-1mm]
&\hspace{15mm}\times
G_{j,\mathsf R}
\left(
\frac{(n+r)\ln 2}{\gamma}
\right).
\label{eq:Gaver_functionals_GWR}
\end{align}
For each fixed HARQ round index \(j\), receiver scheme
\(\mathsf R\), and inversion point \(\gamma\), a separate Wynn-rho
table is constructed. For notational simplicity, the dependence of
\(\rho_{\ell}^{(n)}\) on \(j\), \(\mathsf R\), and \(\gamma\) is
suppressed. These functionals are then accelerated using Wynn's rho
recursion,
\begin{align}
\rho_{-1}^{(n)}
&=0,
\qquad
\rho_{0}^{(n)}
=
\Phi_{n+1}^{(j,\mathsf R)}(\gamma),
\label{eq:Wynn_rho_initialization}\\
\rho_{\ell}^{(n)}
&=
\rho_{\ell-2}^{(n+1)}
+
\frac{\ell}{
\rho_{\ell-1}^{(n+1)}
-
\rho_{\ell-1}^{(n)}
},
\label{eq:Wynn_rho_recursion}
\end{align}
for \(\ell=1,\ldots,M-1\) and
\(n=0,\ldots,M-\ell-1\). An odd value of \(M\) is used so that
\(\rho_{M-1}^{(0)}\) belongs to an even-indexed rho column. For the
fixed \(j\), \(\mathsf R\), and \(\gamma\), the accumulated-SIR CDF is
then approximated by
\begin{equation}
\label{eq:CDF_accumulated_GWR}
F_{\Gamma_{u,\mathsf R}^{(q)}(j)}(\gamma)
=
\mathcal L^{-1}
\left\{
G_{j,\mathsf R}(s)
\right\}(\gamma)
\approx
\rho_{M-1}^{(0)} .
\end{equation}
At the decoding threshold
\(\gamma_{\rm th}=2^{R_0}-1\), the HARQ-CC outage probability after at
most \(C\) rounds is therefore
\begin{equation}
\label{eq:Pout_GWR_common_final}
P_{\mathrm{out}}^{\mathsf R}
=
F_{\Gamma_{u,\mathsf R}^{(q)}(C)}
(\gamma_{\rm th})
\approx
\left.
\rho_{M-1}^{(0)}
\right|_{j=C,\,\gamma=\gamma_{\rm th}},
\end{equation}
where the transform samples entering the rho recursion are computed
from \eqref{eq:G_j_common_abscissa_samples}. 
For the FAS receiver, either the validity-corrected Marcum-$Q$-kernel CDF
\(F_{m,\mathrm{mq}}^{\mathrm{FAS}}\) or the validity-corrected threshold
CDF \(F_{m,\mathrm{th}}^{\mathrm{FAS}}\) is substituted into
\eqref{eq:LST_common_abscissa_GLQ}. The former retains the \(r\)-domain
Gauss--Laguerre quadrature and applies the endpoint normalization in
\eqref{eq:FSIR_MQ_corrected}. The latter obtains its lower complexity
from the raw step threshold before applying the endpoint and monotonicity
corrections in~\eqref{eq:FSIR_threshold_corrected}. Neither raw curve is
inverted. Both admissible CDF approximations are processed using the same
GWR procedure.

\subsection{Average Transmissions and Payload Throughput}
The average number of transmissions and payload throughput are,
respectively,
\begin{equation}
\label{eq:average_transmissions_GWR}
\overline C^{\mathsf R}
=
1+
\sum_{j=1}^{C-1}
F_{\Gamma_{u,\mathsf R}^{(q)}(j)}
(\gamma_{\rm th}),
\end{equation}
and
\begin{equation}
\label{eq:payload_throughput_GWR}
T_{\mathrm{pay}}^{\mathsf R}
=
\frac{
R_0\left(1-P_{\mathrm{out}}^{\mathsf R}\right)
}{
\overline C^{\mathsf R}
}.
\end{equation}

\section{Numerical Results}
\label{subsec:numerical_setup}
This section presents numerical and Monte Carlo (MC) results for the considered HARQ-CC-aided \emph{s}FAMA system. 
Unless otherwise stated, we consider $U=8$ UEs sharing the same
time--frequency resource, use $K=128$ ports for the FAS receiver, and set
the maximum number of HARQ transmissions to $C=4$. Each potential
interfering UE is independently active in each HARQ round with probability
$p_{\rm a}=0.5$. When varying $U$, $p_{\rm a}$ is held fixed, so the resulting performance variation is attributed solely to the change in the number of potential interferers. 
Each UE is equipped with a 1D-FAS of aperture \(15\) cm. The carrier frequency is set to \(7\) GHz, corresponding to the normalized FAS size \(W=3.50\). 
The BR-AS and FPA receivers are included as receiver benchmarks. The endpoint-corrected Marcum-Q-kernel results obtained via SIR-domain convolution in \cite{HARQsFAMA}, along with the corresponding LST-domain results, are included as analytical benchmarks.
The MC results are obtained from the full Jakes' spatial correlation model, whereas the analytical curves are based on the proposed high-correlation spatial block correlation approximation. The block partition uses nearly equal block sizes, with \(B\) set to the number of eigenvalues of the full Jakes correlation matrix exceeding \(\rho_{\mathrm{th}}=1\). The block sizes differ by at most one, and \(\mu^2\) is matched to the nearest neighbor correlation of the full Jakes model. Since the analytical expressions are derived under the high-correlation approximation, they are expected to be most accurate in dense-port settings where \(\mu\) is close to one. 
Table~\ref{tab:approx_validity} summarizes five representative dense-port configurations considered in this work, all of which satisfy \(\mu>0.97\). As \(K\) increases over the fixed aperture, \(\mu\) progressively approaches unity, while the resulting partition consists of eight or nine nearly balanced correlation blocks.
\begin{table}[h!]
\caption{Representative dense-port high-correlation configurations and
the corresponding block-partition parameters.}
\label{tab:approx_validity}
\centering
\begin{tabular}{cccc}
\hline
\(K\) & \(B\) & \(\{L_b\}_{b=1}^{B}\) & \(\mu\) \\
\hline
48  & 8 & \(6,6,6,6,6,6,6,6\)
        & 0.972594 \\
64  & 9 & \(8,7,7,7,7,7,7,7,7\)
        & 0.984748 \\
96  & 9 & \(11,11,11,11,11,11,10,10,10\)
        & 0.993292 \\
128 & 9 & \(15,15,14,14,14,14,14,14,14\)
        & 0.996247 \\
256 & 9 & \(29,29,29,29,28,28,28,28,28\)
        & 0.999069 \\
\hline
\end{tabular}
\end{table}

The HARQ-CC outage probability is evaluated using the SIR-domain
Stieltjes-convolution method in~\eqref{eq:Pout_system_def} and the
LST-domain method in~\eqref{eq:LST_common_abscissa_GLQ}. The former
uses a uniform grid with \(N_{\mathrm{bin}}=6000\) intervals. For the
latter, all receiver schemes use the same common-abscissa construction
and GWR inversion, with
\(N_{\gamma}=256\), \(N_r=30\), \(N_{\tilde r}=30\), and \(M=13\).
The Gaver sums and Wynn-\(\rho\) recursion are evaluated using
50-digit variable-precision arithmetic.

For each activity state \(m\), the Marcum-\(Q\)-kernel endpoint
residual is evaluated using the quadrature expression in
\eqref{eq:MQ_endpoint_residual}, with the same \(N_r\)-point
\(r\)-domain rule as in the raw-CDF evaluation. The corrected CDF in
\eqref{eq:FSIR_MQ_corrected} is then used for activity averaging,
Stieltjes convolution, and LST evaluation.
For the threshold route, the CDF arguments required by the convolution
and LST routines are augmented by a low-SIR grid with
\(4096\) logarithmically spaced points per decade over
\([10^{-14},10^{-1}]\). 
After sorting the combined grid as
\(0=\gamma_0<\gamma_1<\cdots<\gamma_J\), the running-supremum
correction in~\eqref{eq:FSIR_threshold_corrected} is implemented by
\begin{equation}
\label{eq:threshold_cummax_numerical}
\begin{aligned}
f_{m,0}&=0,\\
f_{m,j}
&=
\max\!\left\{
f_{m,j-1},
\overline F_{m,\mathrm{th}}^{\mathrm{FAS}}(\gamma_j)
\right\},
\quad j=1,\ldots,J .
\end{aligned}
\end{equation}
All subsequent threshold-CDF values are obtained by piecewise-linear
interpolation of the corrected table
\(\{(\gamma_j,f_{m,j})\}_{j=0}^{J}\), which covers the required SIR
range. A tolerance of \(\epsilon_{\mathrm{mon}}=10^{-12}\) is used
only to verify the CDF endpoint, range, and monotonicity.
Across all configurations and activity states examined in
Table~\ref{tab:approx_validity}, the largest endpoint residual for the
Marcum-\(Q\)-kernel route is \(2.32\times10^{-6}\), attained at
\(K=48\) and \(m=7\). For the threshold route, the largest endpoint
residual and the largest observed pointwise difference between the
endpoint-extended raw and corrected curves are both
\(2.40\times10^{-12}\), attained at \(K=48\), with the former being
independent of \(m\). The largest SIR location of an observed
low-SIR local minimum is approximately
\(\gamma=7.8\times10^{-3}\) (\(-21.1\)~dB), occurring at \(K=48\)
and \(m=1\). These results indicate that the CDF-validity corrections
are numerically small and that the observed threshold nonmonotonicity
is confined to the extreme low-SIR region. All subsequent performance
evaluations use the corresponding corrected CDFs.

\begin{figure}[tbp]
\begin{center}
\subfloat[Outage probability]{\includegraphics[width = .95\linewidth]{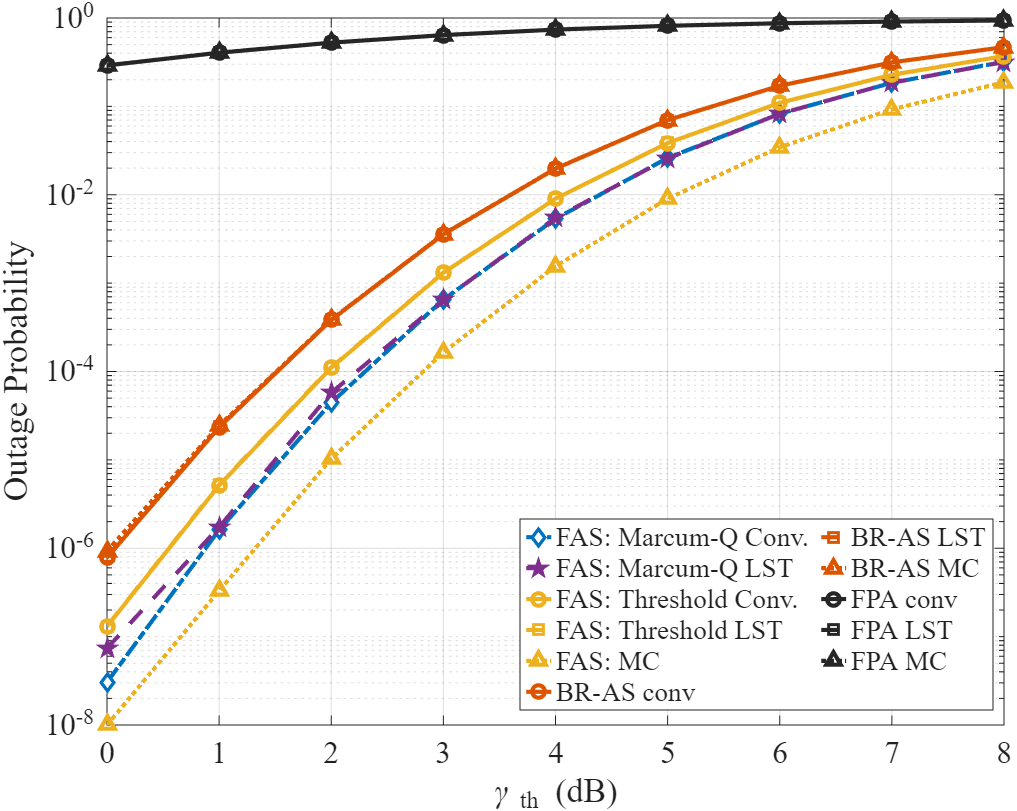}}\\
\vspace{-2mm}
\subfloat[Average Transmissions]{\includegraphics[width = .95\linewidth]{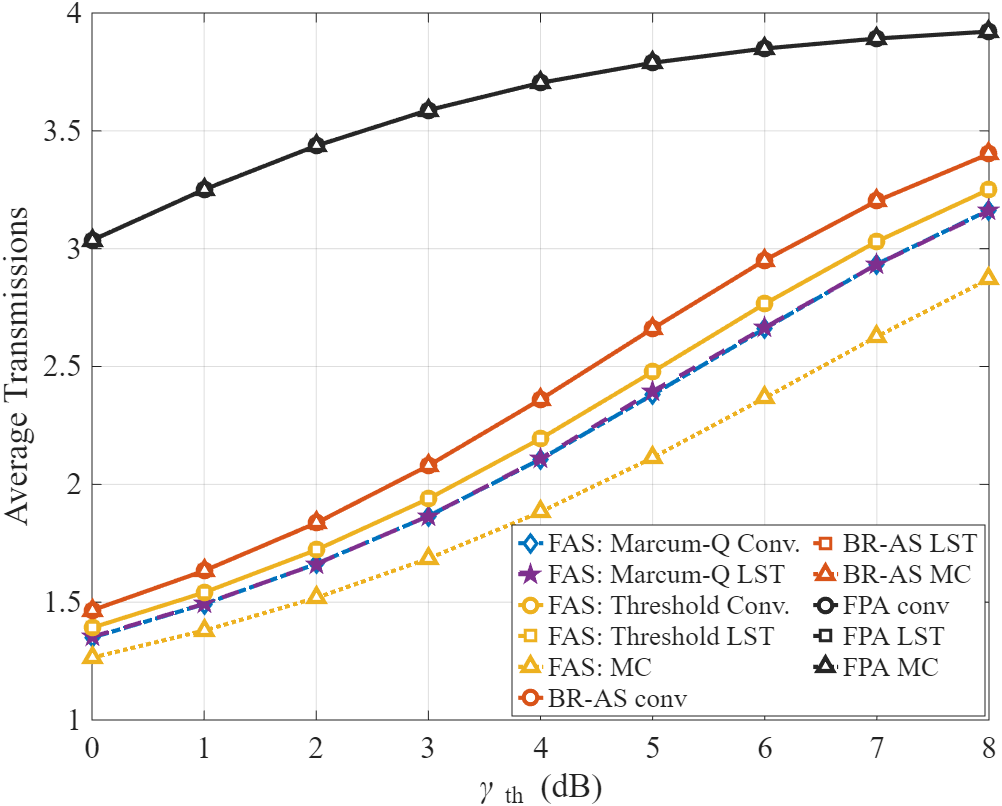}}\\
\vspace{-2mm}
\subfloat[Payload Throughput]{\includegraphics[width = .95\linewidth]{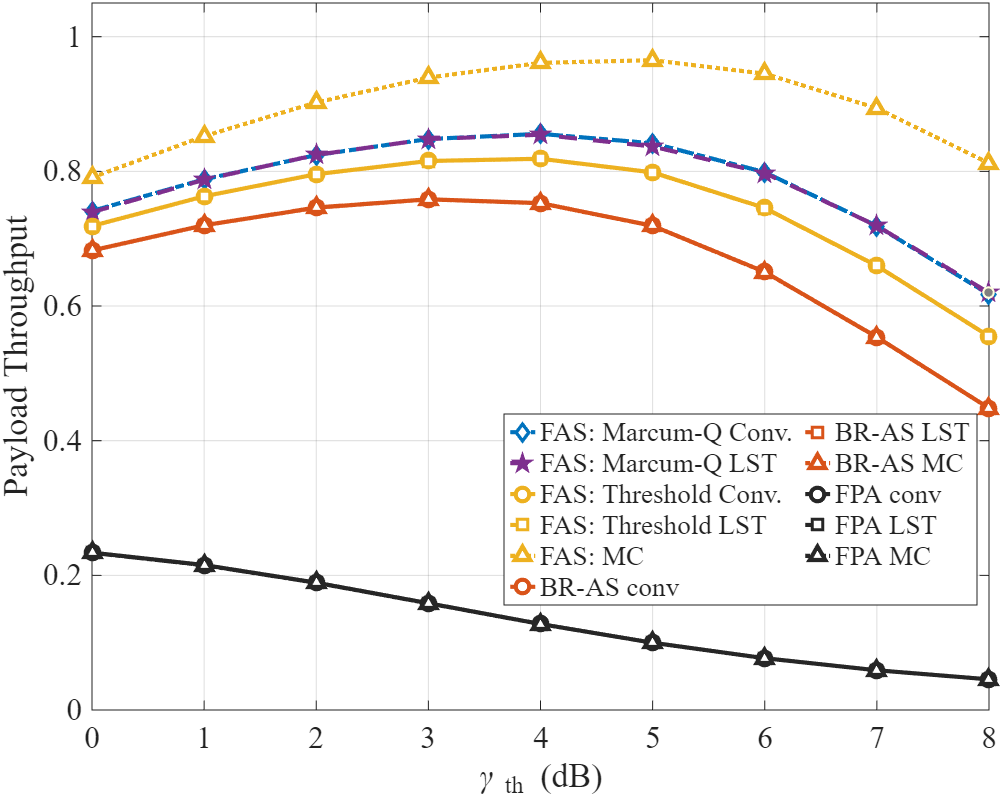}}
\caption{Comparison of the convolution- and LST-domain evaluations with MC simulations for the FAS receiver, its matched BR-AS benchmark, and the FPA receiver versus the SIR decoding threshold $\gamma_{\mathrm{th}}$: (a) outage probability, (b) average number of transmissions, and (c) payload throughput. The Marcum-(Q)-kernel and step-threshold FAS routes use the validity-corrected CDFs in \eqref{eq:FSIR_MQ_corrected} and \eqref{eq:FSIR_threshold_corrected}, respectively.}
\label{fig:performance_validation}
\vspace{-2mm}
\end{center}
\end{figure}

Fig.~\ref{fig:performance_validation} compares the results of SIR-domain convolution and LST-domain evaluation with MC simulations for the three considered receiver schemes. 
Across all three metrics, the SIR-domain and LST-domain results are virtually indistinguishable for each FAS route, confirming the numerical consistency of the two accumulation procedures. This agreement does not, however, by itself establish the accuracy of the underlying high-correlation approximations, since both procedures use the same route-specific validity-corrected per-round CDF in \eqref{eq:FSIR_MQ_corrected} and \eqref{eq:FSIR_threshold_corrected}, respectively. The BR-AS and FPA predictions closely match the corresponding MC results because their per-round SIR distributions do not require the approximations introduced for selected-port FAS analysis. For FAS, the validity-corrected Marcum-(Q)-kernel approximation captures the dependence on \(\gamma_{\mathrm{th}}\) and the relative performance trends, although it yields conservative predictions. The observed discrepancy is consistent with the combined effects of the block-correlation approximation and the omission of the finite remainder $\mathcal R_m$, rather than with the transform-domain evaluation of HARQ combining. The threshold route introduces an additional step-function approximation and is therefore more conservative in the tested cases, while preserving the overall trends.
As \(\gamma_{\mathrm{th}}\)
increases, both the outage probability and the average number of
transmissions increase. The payload throughput initially benefits from
the higher payload rate associated with a larger SIR decoding threshold,
but eventually decreases when the corresponding reliability loss and
retransmission overhead become dominant. %In the tested cases, the MC curves preserve the same receiver ordering and show equal or larger FAS gains over the considered range.
Consequently, the conservative FAS analysis does not exaggerate its advantage over the two benchmarks, as the MC results show even larger performance gains over the considered range.

\begin{figure}[tbp]
\begin{center}
\subfloat[Outage probability]{\includegraphics[width = .95\linewidth]{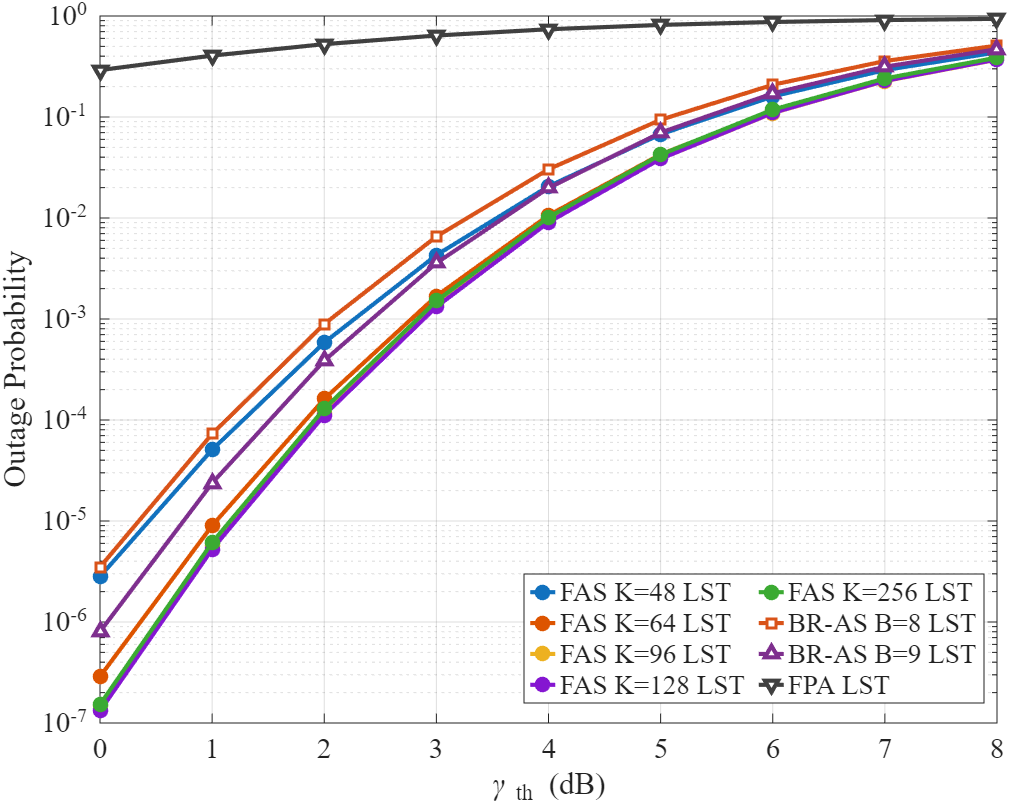}}\\
\vspace{-2mm}
\subfloat[Average Transmissions]{\includegraphics[width = .95\linewidth]{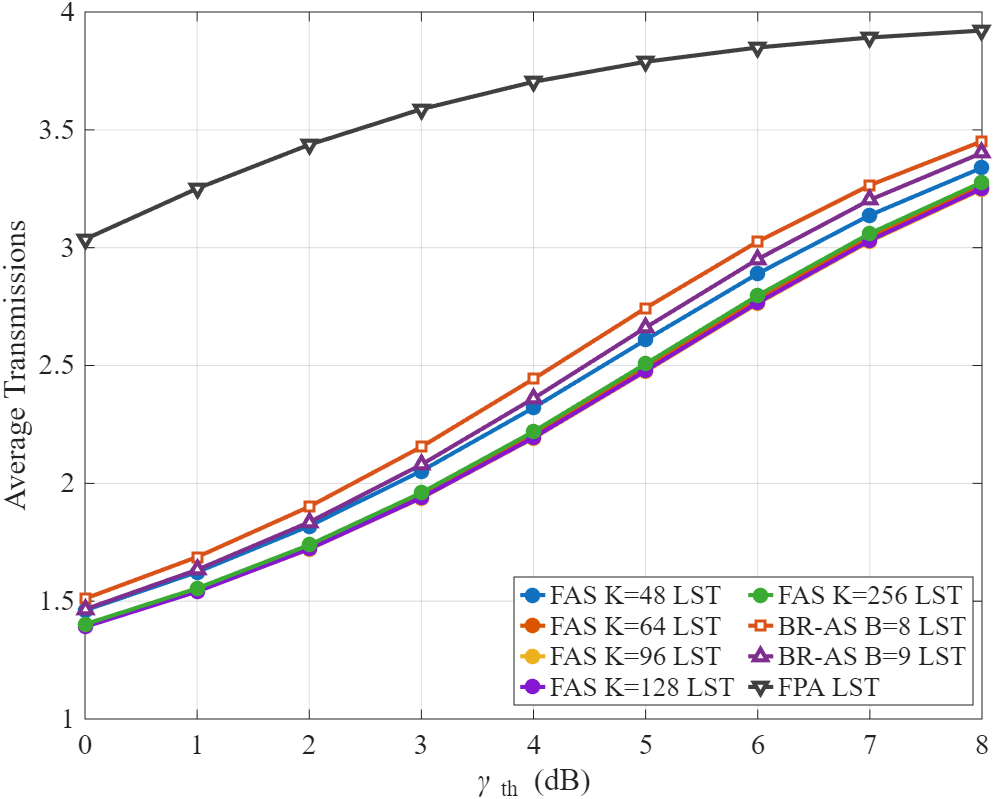}}\\
\vspace{-2mm}
\subfloat[Payload Throughput]{\includegraphics[width = .95\linewidth]{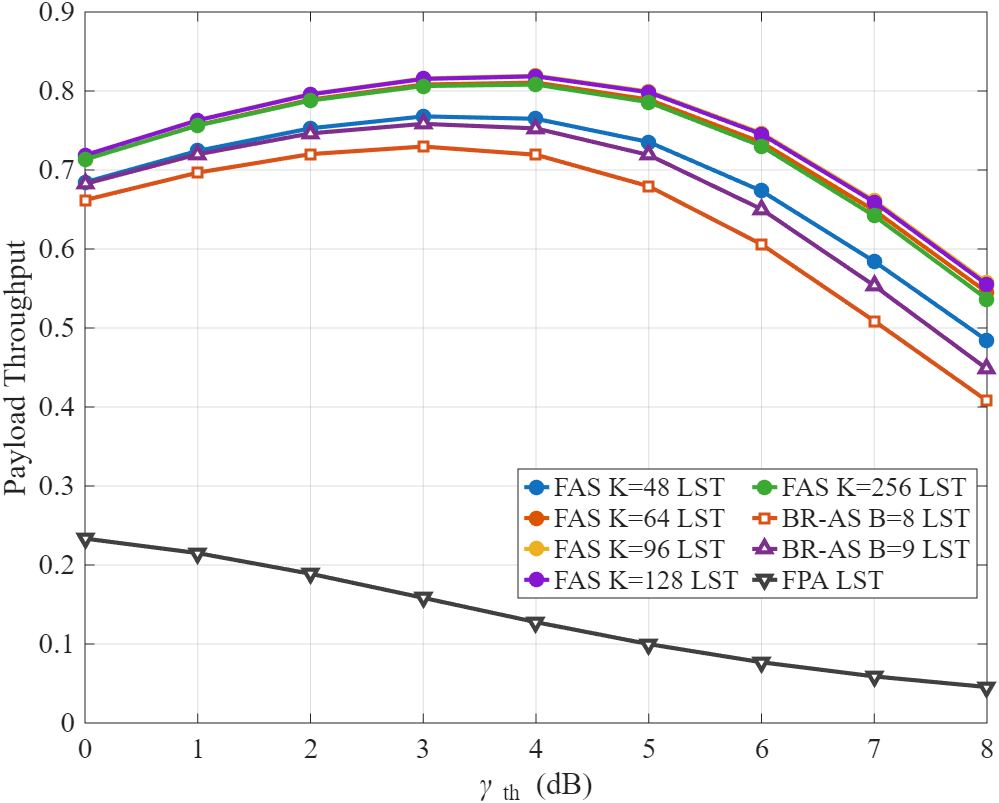}}
\caption{LST-based analytical performance of the FAS threshold route, along with its matched BR-AS benchmark and the FPA benchmark, versus the SIR decoding threshold $\gamma_{\mathrm{th}}$ for different numbers of FAS ports $K$: (a) outage probability, (b) average number of transmissions, and (c) payload throughput.}
\label{fig:performance_threshold_K}
\vspace{-2mm}
\end{center}
\end{figure}

Having established the agreement between the convolution and LST
evaluations, we employ the validity-corrected threshold CDF with the LST
method in Fig.~\ref{fig:performance_threshold_K} to examine the impact of
the number of FAS ports over the considered SIR decoding threshold range.
For $K=48$, the FAS receiver consistently outperforms its corresponding BR-AS benchmark with $B=8$, achieving a lower outage probability and fewer average transmissions, and consequently a higher payload throughput over the entire threshold range. Denser FAS configurations generally provide further performance improvement; however, the curves for $K=64$, $96$, $128$, and $256$ remain closely clustered, suggesting that the performance gain nearly saturates for $K\geq64$. This saturation reflects the diminishing returns of increasing the port density within a fixed aperture, as the additional ports become strongly correlated and provide only limited additional effective selection diversity. In contrast, the FPA receiver exhibits a substantially higher outage probability and requires nearly the maximum number of transmissions at moderate and high SIR decoding thresholds, resulting in a markedly lower payload throughput.

\begin{figure}[tbp]
\begin{center}
\subfloat[Outage probability]{\includegraphics[width = .95\linewidth]{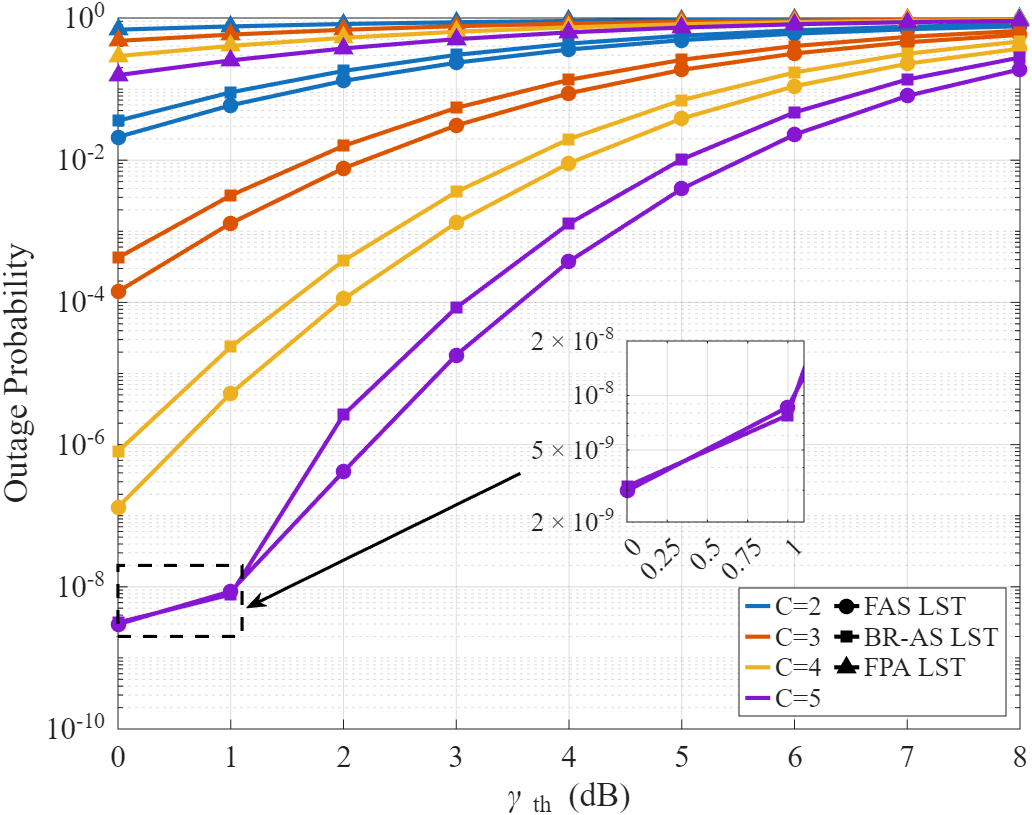}}\\
\vspace{-2mm}
\subfloat[Average Transmissions]{\includegraphics[width = .95\linewidth]{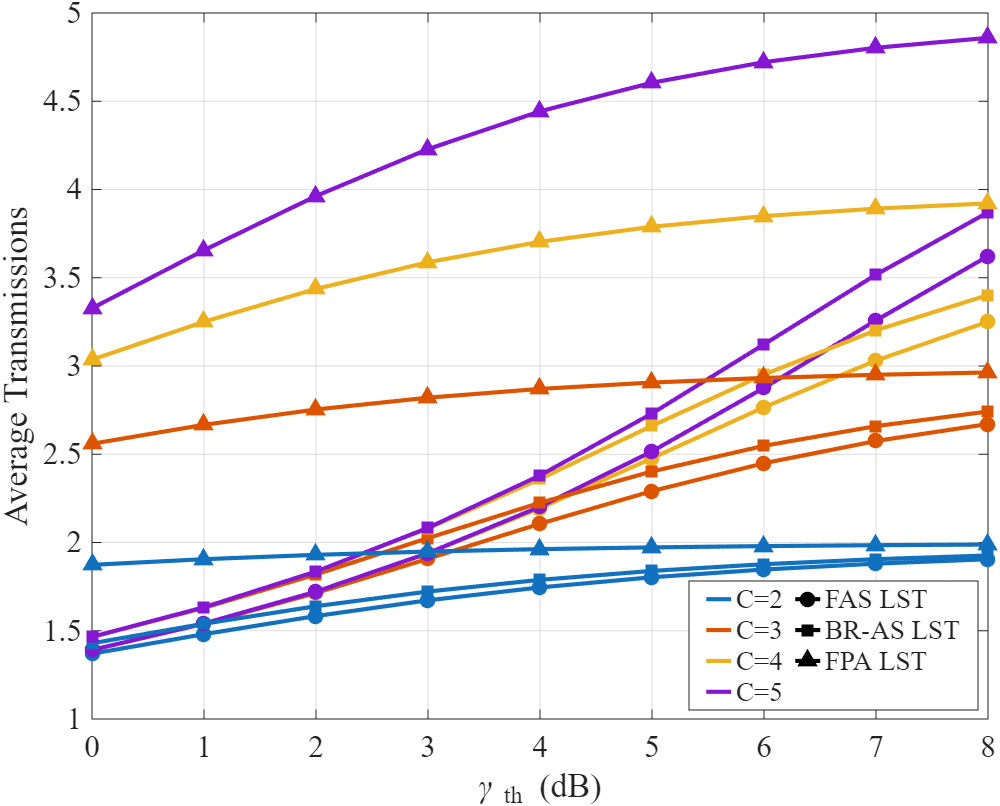}}\\
\vspace{-2mm}
\subfloat[Payload Throughput]{\includegraphics[width = .95\linewidth]{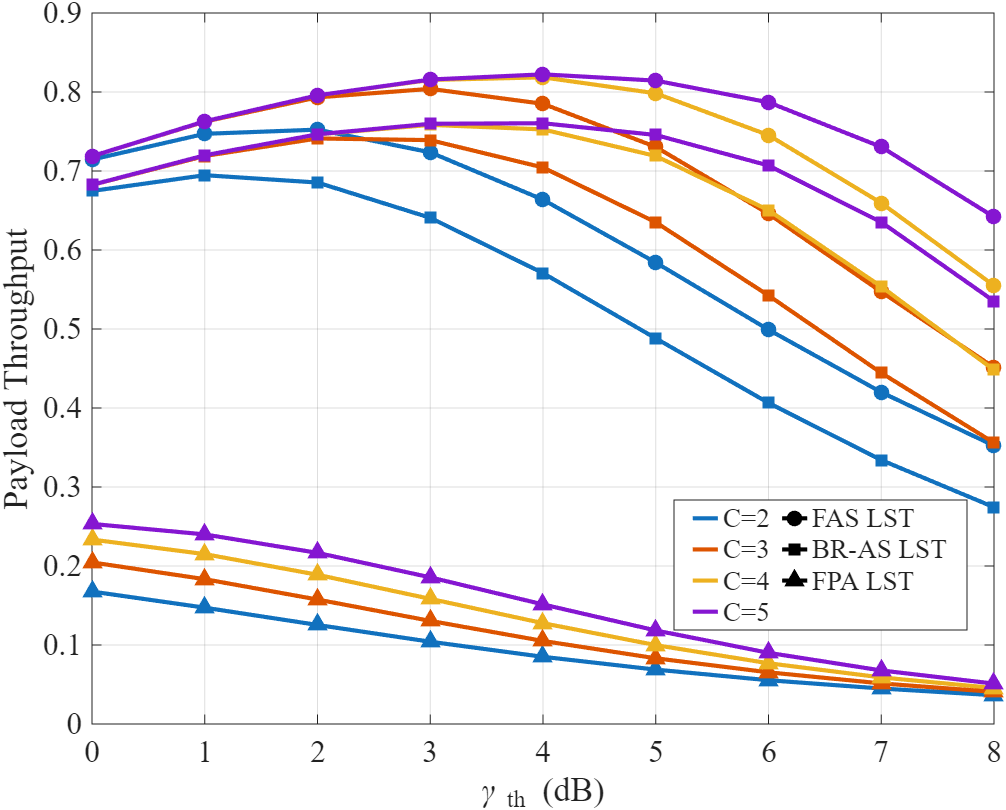}}
\caption{LST-based analytical performance of the FAS threshold route, along with its matched BR-AS benchmark and the FPA benchmark, versus the SIR decoding threshold $\gamma_{\mathrm{th}}$ for different HARQ transmission limits $C$: (a) outage probability, (b) average number of transmissions, and (c) payload throughput.}
\label{fig:performance_threshold_C}
\vspace{-2mm}
\end{center}
\end{figure}

Fig.~\ref{fig:performance_threshold_C} examines the impact of the maximum number of HARQ transmissions $C$. Increasing $C$ allows additional chase-combining rounds, substantially reducing the outage probability of the FAS and BR-AS receivers, particularly at low and moderate SIR decoding thresholds. This reliability improvement comes at the expense of a higher average number of transmissions, with the increase becoming more pronounced at larger $\gamma_{\mathrm{th}}$, where early decoding is less likely. For the FAS receiver, the outage reduction outweighs the additional retransmission overhead, resulting in a higher payload throughput as $C$ increases. The widening gap between the $C=4$ and $C=5$ throughput curves at larger $\gamma_{\mathrm{th}}$ further shows that additional HARQ rounds become more beneficial under stringent decoding requirements. Although BR-AS follows the same general trend, FAS consistently achieves lower outage, fewer transmissions, and higher throughput by exploiting the port-level channel variations within each block. In contrast, the FPA receiver operates close to the transmission limit and retains a high outage probability over much of the considered range, so increasing $C$ provides only a modest throughput improvement.

\begin{figure}[tbp]
\begin{center}
\subfloat[Outage probability]{\includegraphics[width = .95\linewidth]{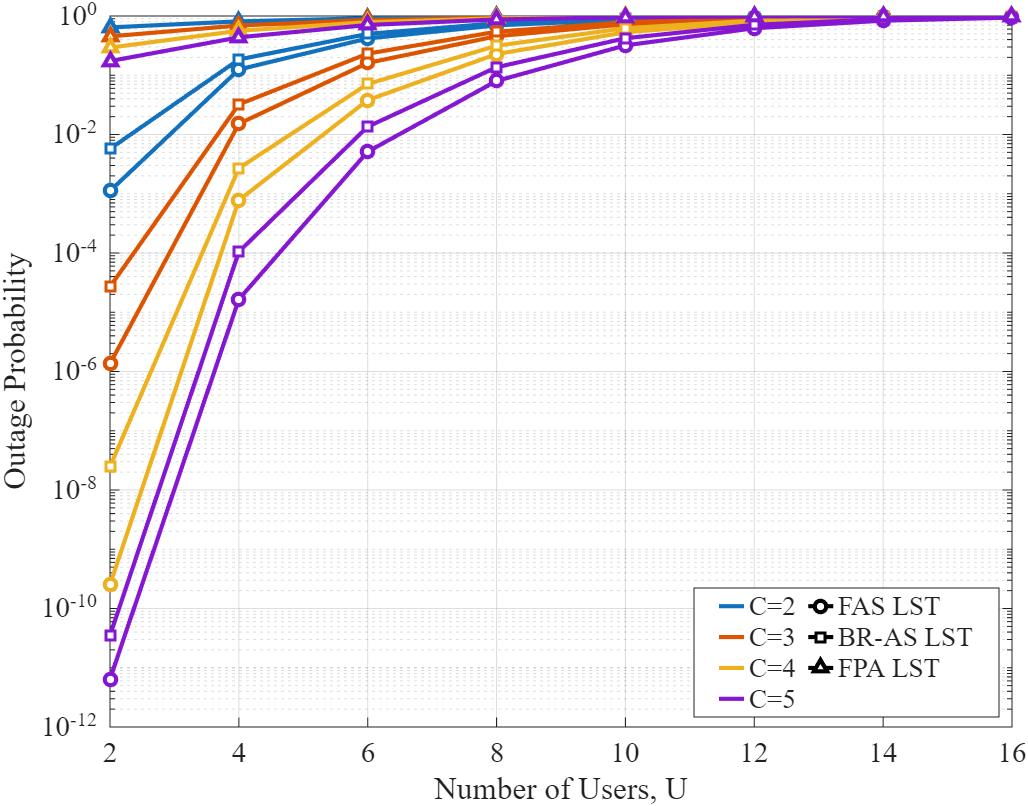}}\\
\vspace{-2mm}
\subfloat[Average Transmissions]{\includegraphics[width = .95\linewidth]{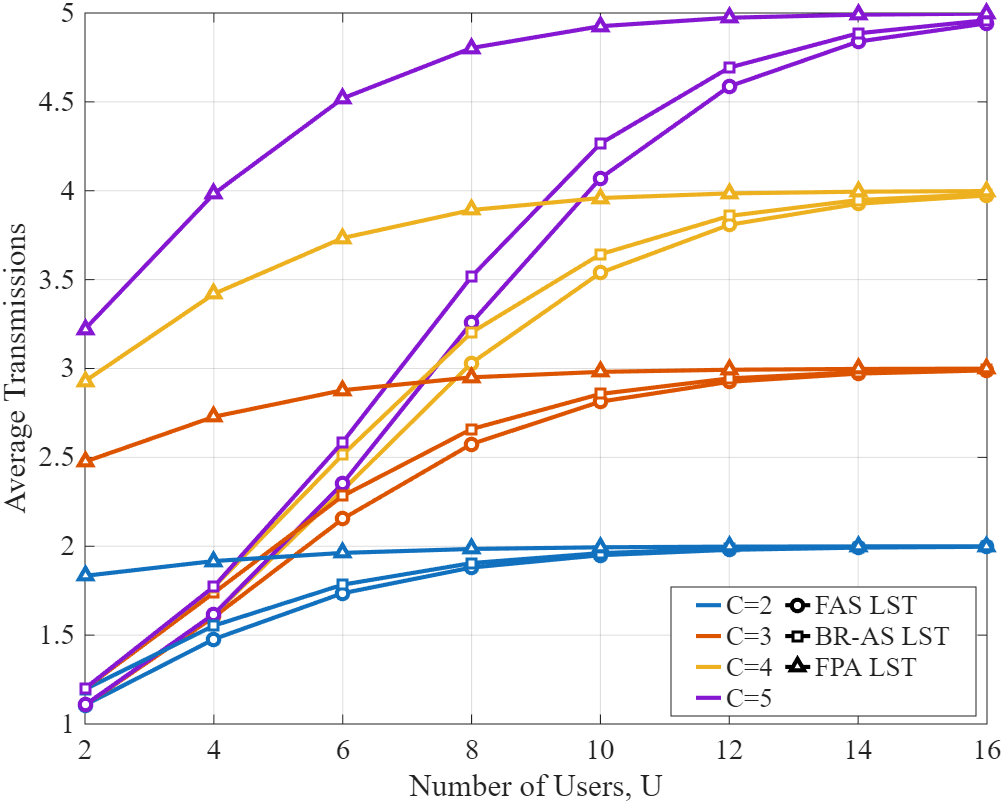}}\\
\vspace{-2mm}
\subfloat[Payload Throughput]{\includegraphics[width = .95\linewidth]{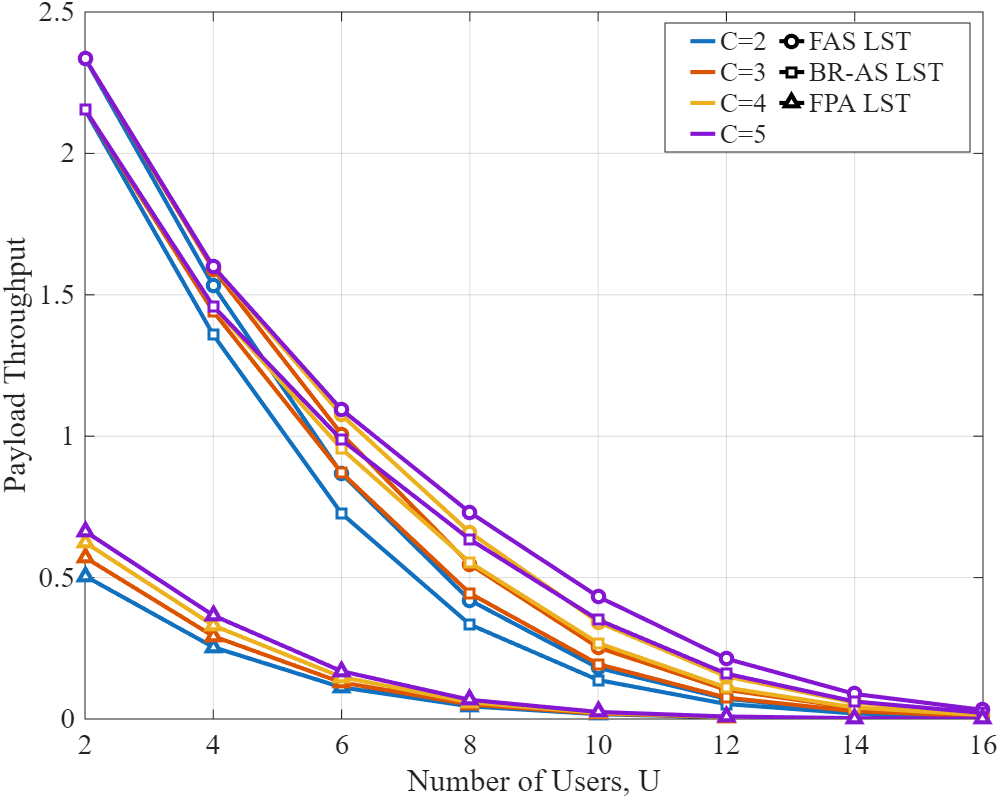}}
\caption{LST-based analytical performance of the FAS threshold route, along with its matched BR-AS benchmark and the FPA benchmark, versus the number of users $U$ for different HARQ transmission limits $C$: (a) outage probability, (b) average number of transmissions, and (c) payload throughput.}
\label{fig:performance_users_C}
\vspace{-2mm}
\end{center}
\end{figure}

Fig.~\ref{fig:performance_users_C} examines the impact of the number of users $U$ under different HARQ transmission limits $C$. 
At fixed $p_{\rm a}$, increasing $U$ increases the average number of active interferers and thus strengthens aggregate cochannel interference. This increases the outage probability and the average number of transmissions, while reducing the per-user payload throughput for all receiver schemes.
For small \(U\), increasing \(C\) markedly reduces the outage probabilities
of the FAS and BR-AS receivers while causing only a
limited increase in the average number of transmissions. The resulting
reliability improvement offsets the additional retransmission overhead,
thereby maintaining or improving the payload throughput. As \(U\) increases, the outage probability approaches one and the average number of transmissions approaches \(C\), indicating that most packets use all available HARQ rounds but remain unsuccessfully decoded. A larger \(C\)
generally yields higher payload throughput over the considered range,
although the gain is negligible when the reliability is already high and
eventually diminishes as the throughput curves converge in the
strong-interference regime. For any fixed \(C\), the FAS receiver generally achieves lower outage probability and average number of transmissions, together with higher payload throughput, than BR-AS. These performance gaps eventually narrow in the large-\(U\) regime as both receivers approach their strong-interference limits. Compared with the other two receivers, the FPA receiver reaches an average number of transmissions close to \(C\) at smaller \(U\), and generally exhibits the highest outage probability and the lowest payload throughput.

\begin{figure}[tbp]
\centering
\includegraphics[width=.95\linewidth]{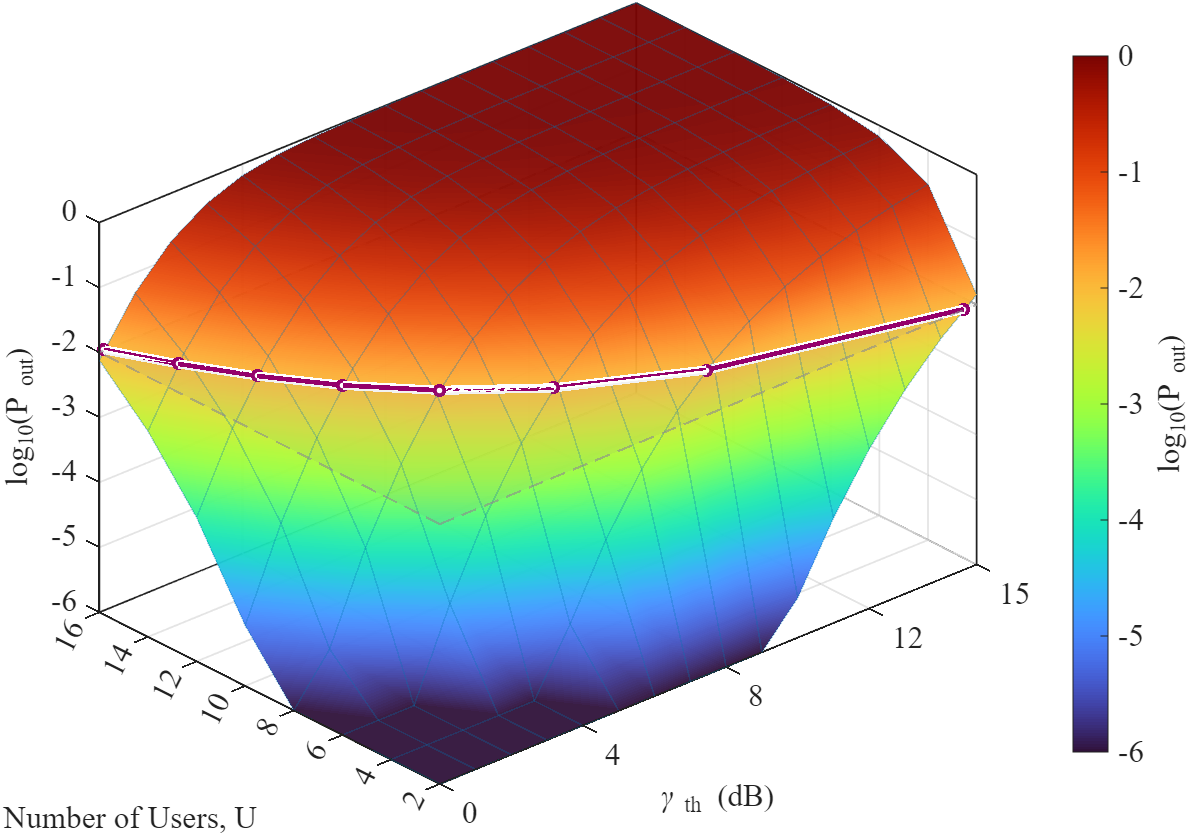}
\caption{LST-based analytical outage performance of the FAS threshold route versus the SIR decoding threshold $\gamma_{\rm th}$ and the number of users $U$. The magenta curve denotes the reliability boundary at $P_{\rm out}=10^{-2}$, while the gray plane represents the corresponding $1\%$ outage level.}
\label{fig:joint_outage_gamma_U}
\vspace{-2mm}
\end{figure}

Fig.~\ref{fig:joint_outage_gamma_U} illustrates the joint dependence of the outage probability of HARQ-\emph{s}FAMA on the SIR decoding threshold $\gamma_{\rm th}$ and the number of users $U$. Over the considered parameter range, the outage probability increases monotonically with both parameters. For a given $U$, a larger $\gamma_{\mathrm{th}}$ imposes a more stringent decoding requirement, whereas for a fixed $\gamma_{\mathrm{th}}$, increasing $U$ raises the expected number of active cochannel interferers under the adopted activity model and hence strengthens the aggregate interference. Consequently, the largest outage probabilities are observed in the high load and high threshold region.
The gray plane represents the target outage level
$P_{\rm out}=10^{-2}$. For each considered $U$, the magenta marker
identifies the numerically estimated maximum admissible SIR decoding
threshold $\gamma_{\rm th}^{\star}(U)$, defined as the largest value of
$\gamma_{\rm th}$ satisfying $P_{\rm out}\leq 10^{-2}$. The markers
are connected only as a visual guide and are plotted slightly above the
target plane for clarity. Since $P_{\rm out}$ increases monotonically
with $\gamma_{\rm th}$, the reliability-feasible threshold range for
a given $U$ is
$\gamma_{\rm th}\leq\gamma_{\rm th}^{\star}(U)$.
As the number of users increases,
$\gamma_{\rm th}^{\star}(U)$ decreases from $14.64$~dB at $U=2$ to
$0.13$~dB at $U=16$, demonstrating a substantial contraction of the
reliability-feasible operating region. The resulting boundary
therefore quantifies the reliability-constrained tradeoff between the
supported user population and the admissible SIR decoding threshold, rather
than identifying a unique optimal operating point.

%\vspace{-2mm}

\section{Conclusion}
\label{Conclusion}
This paper investigated downlink HARQ-CC-aided \emph{s}FAMA in interference-limited networks with per-round SIR-maximizing port reselection and chase combining across independent fading blocks. Using the spatial block correlation model, we formulated a validity-corrected high-correlation Marcum-$Q$-kernel approximation and a lower-complexity step-threshold approximation for the selected-port SIR distribution. Closed-form counterparts were obtained for BR-AS and FPA. When driven by
the same corrected per-round CDF, the convolution- and LST-domain
implementations produced consistent results, while MC simulations
corroborated the predicted trends and receiver ordering over the tested
configurations. Over the tested configurations, the results showed the
performance advantage of the FAS receiver over the considered benchmarks
and indicated that maintaining
a prescribed outage level requires a lower decoding threshold as the user
load increases. They also revealed that FAS performance gradually
saturates with increasing port density, while the throughput benefit of
additional HARQ rounds becomes marginal under severe multiuser
interference.

\appendices
\section{Derivation of the Raw Transition Threshold}
\label{sec:threshold_derivation}
It remains to calculate the raw transition threshold \(\delta_{m,b}(\tilde r_{u,b};\gamma)\). For each CDF evaluation at a given \(\gamma>0\), consider a fixed value of the outer integration variable \(\tilde r_{u,b}\), and define
\begin{equation}
\label{eq:alpha_z_def}
\alpha
\triangleq
\chi_\mu
\sqrt{\frac{2\gamma \tilde r_{u,b}}{\gamma+1}},
\qquad
z
\triangleq
\chi_\mu
\sqrt{\frac{2 r_{u,b}}{\gamma+1}} .
\end{equation}
By~\eqref{eq:mq_kernel_definition}, these variables give
$\mathcal Q_m(\gamma;r_{u,b},\tilde r_{u,b})=Q_m(\alpha,z)$.
For fixed \((\gamma,\tilde r_{u,b})\), \(\alpha\) is constant, whereas \(z\) increases monotonically with \(r_{u,b}\). Since $Q_m(\alpha,z)$ is monotonically decreasing in $z$, the sharp transition of $[Q_m(\alpha,z)]^{L_b}$ in the dense-port regime can be characterized by its maximum-descent point. The derivative is
\begin{equation}
\label{eq:dQdz_exact}
\begin{aligned}
\frac{\partial [Q_m(\alpha,z)]^{L_b}}{\partial z}
&=
-L_b[Q_m(\alpha,z)]^{L_b-1}
\frac{z^{m}}{\alpha^{m-1}}  \\
&\quad \times
\exp\!\left(-\frac{z^{2}+\alpha^{2}}{2}\right)
I_{m-1}(\alpha z),
\end{aligned}
\end{equation}
where $I_{m-1}(\cdot)$ is the modified Bessel function of the first kind.
For large $\alpha z$, using
$I_{m-1}(\alpha z)\approx (2\pi\alpha z)^{-1/2}e^{\alpha z}$ gives
\begin{equation}
\label{eq:dQdz_asymp}
\begin{aligned}
\frac{\partial [Q_m(\alpha,z)]^{L_b}}{\partial z}
&\approx
-\frac{L_b}{\sqrt{2\pi}}
[Q_m(\alpha,z)]^{L_b-1}
\frac{z^{m-\frac{1}{2}}}{\alpha^{m-\frac{1}{2}}} \\
&\quad \times
\exp\!\left(-\frac{(\alpha-z)^2}{2}\right).
\end{aligned}
\end{equation}
This large-argument Bessel approximation is non-uniform as
\(\gamma\downarrow0\), because \(\alpha\to0\) for fixed
\(\tilde r_{u,b}\). The threshold derived below therefore defines the raw
step approximation; its endpoint and monotonicity are treated separately
by~\eqref{eq:threshold_endpoint_normalized}--\eqref{eq:FSIR_threshold_corrected}.
The transition occurs around $z=\alpha$. A first-order approximation of the
maximum-descent point around this transition yields
\begin{equation}\label{eq:zstar_closed_form_2}
z_{m,b}^\star(\alpha)
\approx
\alpha
+
\frac{
\Big(m-\frac{1}{2}\Big)-\alpha\,\kappa_b
}{
\kappa_b\Big(m-\frac{1}{2}\Big)+\alpha
},
\qquad
\kappa_b\triangleq \sqrt{\frac{L_b-1}{2\pi}}.
\end{equation}
Finally, the threshold $\delta_{m,b}(\tilde r_{u,b};\gamma)$ is obtained by mapping $z_{m,b}^\star$ back to
$r_{u,b}$ via~\eqref{eq:alpha_z_def}, i.e.,
\begin{equation}
\label{eq:delta_def_map}
\delta_{m,b}(\tilde r_{u,b};\gamma)
\triangleq
\frac{\gamma+1}{2\chi_\mu^2}
\left[
z_{m,b}^{\star}(\alpha)
\right]^2.
\end{equation}
Substituting~\eqref{eq:zstar_closed_form_2} and~\eqref{eq:alpha_z_def}
into~\eqref{eq:delta_def_map} gives the following closed-form threshold
approximation:
\begin{equation}
\label{eq:delta_threshold_appendix}
\begin{aligned}
\delta_{m,b}(\tilde r_{u,b};\gamma)
&=
\left(
\sqrt{\gamma \tilde r_{u,b}}
+
\beta
\frac{
\left(m-\frac{1}{2}\right)\beta
-\kappa_b\sqrt{\gamma \tilde r_{u,b}}
}{
\kappa_b\left(m-\frac{1}{2}\right)\beta
+\sqrt{\gamma \tilde r_{u,b}}
}
\right)^{2},
\end{aligned}
\end{equation}
where $\beta=\chi_\mu^{-1}\sqrt{(\gamma+1)/2}$.
This raw threshold is used in the step-function approximation of the
per-block selection event. Before either Stieltjes convolution or LST
evaluation, the resulting raw CDF is converted into
\(F_{m,\mathrm{th}}^{\text{FAS}}\) according to
\eqref{eq:threshold_endpoint_normalized}--\eqref{eq:FSIR_threshold_corrected}.

\bibliographystyle{IEEEtran}

\end{document}